\documentclass[12pt,showpacs]{iopart}
\usepackage{graphicx,color}
\usepackage{bm}
\usepackage{fleqn}
\usepackage{amssymb}
\usepackage{epstopdf}

\newcommand{\cth}{\mathop{\rm cth}\nolimits}

\usepackage[normalem]{ulem}

\begin{document}

\title{Thermal and quantum fluctuations in chains of ultracold polar molecules} \author{G. E. Astrakharchik$^1$, Gabriele~De~Chiara$^2$, Giovanna Morigi$^2$, and Jordi Boronat$^1$}
\address{$^1$ Departament de F\'{\i}sica i Enginyeria Nuclear, Campus Nord B4-B5, Universitat Polit\`ecnica de Catalunya, E-08034 Barcelona, Spain} \address{$^2$ Grup d'\`Optica, Departament de F\'{\i}sica, Universitat Aut\`onoma de Barcelona, 08193 Bellaterra, Spain}

\date{\today} \begin{abstract}
Ultracold polar molecules, in highly anisotropic traps and interacting via a repulsive dipolar potential, may form one-dimensional chains at high densities. According to classical theory, at low temperatures there exists a critical value of the density at which a second order phase transition from a linear to a zigzag chain occurs. We study the effect of thermal and quantum fluctuations on these self-organized structures using classical and quantum Monte Carlo methods, by means of which we evaluate the pair correlation function and the static structure factor. Depending on the parameters, these functions exhibit properties typical of a crystalline or of a liquid system. We compare the thermal and the quantum results, identifying analogies and differences. Finally, we discuss experimental parameter regimes where the effects of quantum fluctuations on the linear - zigzag transition can be observed.
\end{abstract}
\maketitle

\section{Introduction}

The experimental realization of strongly interacting atomic ensembles at ultralow temperatures allows the observation of self-organization into ordered patterns. Prominent examples are Coulomb crystals of laser-cooled ions in Paul and Penning traps~\cite{Dubin99}. In these systems laser cooling permits one to access the crystalline regime, which emerges from the competition of the long-range Coulomb interaction and the trapping potential at low temperature. Different structures are found depending on the trap geometry and the number of atoms \cite{Dubin99,Birkl92,Waki92,Drewsen08}. In one of the early experimental works~\cite{Birkl92,Waki92}, the phase diagram of the various low dimensional crystalline structures was determined as a function of the trapping frequency of the harmonic confinement and of the number of atoms, starting from a setup with the ions arranged in a linear chain. Later works, concerned about the possibility to measure quantum fluctuations affecting the structure of these systems, pointed out that the ground state of ion crystals is essentially classical~\cite{Javanainen95}. Recent theoretical works, nevertheless, predicted that for chains of few ions quantum fluctuations, such as single-particle tunneling at the structural transition linear -- zigzag, may be observed~\cite{Retzker08}.

Under this perspective, ultracold dipolar gases of atoms or polar molecules are interesting systems, as the nature of the dipolar interaction allows one to observe the interplay between quantum degeneracy and long-range forces. Dipolar gases of ultracold atoms have been experimentally realized with Chromium atoms~\cite{Pfau07,Pfau08}. In these experiments the strength of the $s$-wave scattering interaction is conveniently controlled by tuning the Feshbach resonance and can be made vanishing, thus leading to the realization of purely dipolar systems. Stability against collapse is warranted by polarizing the gas in a two-dimensional geometry, such that the dipolar interactions can be considered as purely repulsive~\cite{buchler07, Astrakharchik07a}. Several theoretical works have studied the phases of dipolar gases of atoms or polar molecules as a function of density and dimensionality~\cite{Goral02,Santos03,Menotti07,Kollath08,Parker08,Nath09}. In the context of this paper, particularly relevant are studies which predicted the creation of self-organized structures in a two-dimensional bulk system, showing that the ground state may exhibit the typical features of a crystal or of a quantum fluid depending on the density~\cite{buchler07, Astrakharchik07a}.

One-dimensional dipolar gases can be experimentally realized in highly anisotropic traps~\cite{Kinoshita05}. In this limit, Luttinger liquid models describe the long-range properties of dipolar gases at ultralow temperatures~\cite{Citro07}. In this case, by tuning the density, the phase of the gas undergoes a crossover from a Tonks--Girardeau gas to a crystal-like phase~\cite{Arkhipov05}. Differing from Coulomb systems, the crystal-like phase is energetically favorable at large densities, as it is typical of short-range interacting systems.

In Ref.~\cite{Astrakharchik08b} we presented a theoretical study of low dimensional dipolar gases focusing on the effects of quantum fluctuations on the formation of transverse correlations at the classical critical point from a linear to a planar distribution, where the one-dimensional structure is unstable with respect to increasing the atomic density and/or to decreasing the transverse potential. In the present article, we provide further details of that study, and extend it to considering thermal effects in the formation of quasi-crystalline structures in the classical domain. The analysis of the thermal and quantum fluctuations is performed numerically, by means of Monte Carlo simulations. The pair correlation function and the structure form factor for the linear and zigzag chains for the classical case at finite temperature and the quantum case at $T=0$ are compared, and analogies and differences discussed.

This article is organized as follows. In Sec.~\ref{Sec:Model} we introduce the theoretical model for a two-dimensional gas of dipolar bosons in presence of tight transverse confinement. In Sec.~\ref{Sec:3} we discuss the mechanical stability of a one-dimensional array of dipoles, assuming classical dynamics and neglecting thermal fluctuations. In Sec.~\ref{Sec:4} the numerical methods applied for simulating the thermal and quantum fluctuations are described. The effect of thermal fluctuations on the quasi-crystalline phase are discussed in Sec.~\ref{Sec:5}. The phase diagram for the quantum ground state is introduced and discussed in Sec.~\ref{Sec:6} while the numerical results for the pair correlation function and structure form factor of linear and zigzag phases are presented in  Sec.~\ref{Sec:7}. The conclusions are drawn in Sec.~\ref{Sec:8}, and details of the calculations are reported in the appendices.

\section{The model} \label{Sec:Model}

We consider $N$ bosonic atoms or molecules of mass $m$ at ultralow temperatures. The particles have induced or permanent dipole and are confined by an external harmonic potential on the plane $x-y$ at $z=0$ by means of a very steep confinement along the $z$ axis, while their dipoles are aligned along the $z$ direction by means of an external field, such that their mutual interaction is always repulsive. By appropriately tuning an external magnetic field to the top of a Feshbach resonance, the $s$-wave scattering can be made negligible so that the particles interact exclusively through the repulsive long-range dipolar interaction potential, which here reads
\begin{equation}
V_{int}=\frac{C_{dd}}{4\pi|{\bm \rho_1}-{\bm \rho_2}|^3},
\end{equation}
where $C_{dd}$ is the dipolar interaction strength and ${\bm \rho_i}=(x_i,y_i)$ is the particle position vector in the $x-y$ plane.

The Hamiltonian of the dipoles contains the kinetic energy, the harmonic confinement and the dipole-dipole interaction
\begin{eqnarray}
\label{Eq:ham} 
H =\sum_j\left [ \frac{  {\bm p_j}^2}{2m} + \frac{1}{2}m\nu_t^2y_j^2\right]
+\frac{1}{2}\sum_{j=1}^N\sum_{j\neq i}\frac{C_{dd}}{4\pi |{\bm \rho_i}-{\bm \rho_j}|^3},
\end{eqnarray}
where ${\bm p_j}$ is the conjugate momentum of the coordinate ${\bm \rho_j}$ and $\nu_t$ is the confinement frequency of the particles along the $y$-direction. In the following we will consider a homogeneous system along the $x$-direction with periodic boundary conditions. For the classical regime, where the particles are distinguishable, this will correspond to imposing ${\bm \rho_{i+N}}={\bm \rho_i}$. In the quantum regime, this consists in imposing that the ground-state wave function fulfills the condition $\psi (x_1,\ldots,x_j+L,\ldots,x_N,y_1\ldots,y_N) = \psi(x_1,\ldots,x_j,\ldots,x_N,y_1,\ldots,y_N)$ for all $j$, where $L$ is the size of the system along the $x$-direction.


\section{Mechanical stability of a classical chain of dipoles}
\label{Sec:3}

In this section the dipolar system is studied in the classical regime. We first consider the equilibrium positions of the dipoles as determined by the trap and the repulsive dipolar potential, and compute the critical value of the density at which a harmonic expansion around these equilibrium positions is unstable.

\begin{figure}
\begin{center}
\includegraphics[width=0.5\columnwidth, angle=-90]{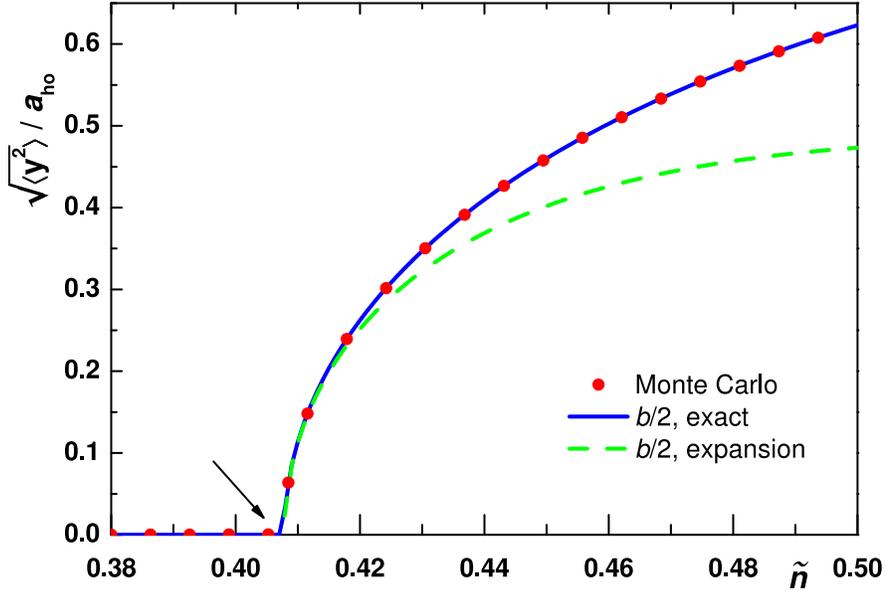}
\caption{(color online)
Transverse variance of the dipolar distribution for the classical case, $\sqrt{\langle y^2\rangle}$, (in units of $a_{ho}=\sqrt{E_{ho}/m\nu_t^2}$) as a function of the linear density $\tilde n$. The variance is evaluated from the classical equilibrium positions of the potential, and it does not contain the thermal fluctuation: it hence gives the displacement of the dipoles from the trap axis at equilibrium. The value of $b$ in classical system can be directly compared to the mean transverse spreading $\sqrt{\langle y^2\rangle}$ as both are expected to coincide in linear and zigzag chains at low temperatures. The critical value $\tilde{n}_c$ where the transition linear-zigzag chain occurs is shown by the arrow. The solid line corresponds to the solution of Eq.~\eref{Eq:b:exact}, the symbols to the results of a classical Monte Carlo simulation, the dashed line is a Taylor expansion of Eq.~\eref{Eq:b:exact}, see Eq.~\eref{Eq:b:Landau}. Here, $\tilde{n}\tilde r_0=3$.
}\label{fig:b}
\end{center}
\end{figure}

For later convenience, we rescale the Hamiltonian in terms of a unit energy for the transverse oscillator, $E_{ho}$, and define the rescaled classical variables ${\bm \pi_j}={\bm p_j}/\sqrt{m E_{ho}}$ and ${\bm \varrho_j}={\bm \rho_j}/\sqrt{E_{ho}/m\nu_t^2}$, where now ${\bm \varrho_j}=(\tilde{x}_j,\tilde{y}_j)$. In these units, the dimensionless Hamiltonian $\tilde H = H/E_{ho}$ reads
\begin{equation} \label{eq:hamtilde}
\tilde{H}=\frac{1}{2}\sum_j{\bm \pi_j}^2+\tilde V \end{equation}
where the potential energy $\tilde V$ is given by
\begin{equation}
\label{eq:tildeV} \tilde V=\frac 12 \sum_j \left[\tilde y_j^2+\sum_{i\neq  j}\frac{\tilde{r}_0}{|{\bm \varrho}_i-{\bm
\varrho}_j|^3}\right], \end{equation}
and $\tilde r_0=r_0/\sqrt{ E_{ho}/m\nu_t^2}$ with \begin{equation} r_0=\frac{m\nu_t^2C_{dd}}{4\pi E_{ho}^2} \end{equation} a characteristic length, whose physical meaning will become more transparent when considering the quantum mechanical case. The classical equilibrium positions of potential~\eref{eq:hamtilde} are here denoted by the vector ${\bm \varrho_i^{(0)}} = (\tilde x_i^{(0)}, \tilde y_i^{(0)})$, which is solution of the equations \begin{eqnarray} \left . \frac{\partial
\tilde V}{\partial \tilde x_i}\right |_{{\bm \varrho_i}={\bm
\varrho_i^{(0)}}}   &=&\left . -3 \tilde r_0\sum_{j\neq
i}\frac{\tilde x_i-\tilde x_j}{|{\bm \varrho_i}-{\bm
\varrho_j}|^5}\right |_{{\bm \varrho_i}={\bm \varrho_i^{(0)}}}=0,
\label{Eq:x}
\\
\left . \frac{\partial \tilde V}{\partial \tilde y_i}\right
|_{{\bm \varrho_i}={\bm \varrho_i^{(0)}}}   &=&\left .  \tilde
y_i-3\tilde r_0\sum_{j\neq i}\frac{\tilde y_i-\tilde y_j}{|{\bm
\varrho}_i-{\bm \varrho_j}|^5}\right |_{{\bm \varrho_i}={\bm
\varrho_i^{(0)}}}=0. \label{Eq:y} \end{eqnarray} We assume the interparticle distance along the $x$ axis to be uniform, and given by the length
\begin{equation} a=1/n\,,\end{equation} where $n$ is the linear density and we denote by $\tilde{n}=n\sqrt{E_{ho}/m\nu_t^2}$ the rescaled density. Within the assumption of equispaced dipoles, Eq.~\eref{Eq:x} is automatically satisfied. In order to find a solution of Eq.~\eref{Eq:y}, we assume the {\it ansatz} ${\bm \varrho_j^{(0)}} = (j\tilde a, (-1)^j \tilde b /2 )$, with $\tilde a=1/\tilde n$, and which for $\tilde b=0$ corresponds to a linear arrangement of the dipoles, while for $\tilde b\neq 0$ it gives a zigzag structure. Using this {\it ansatz}, Eq.~\eref{Eq:y} becomes
\begin{equation}
\label{Eq:b:exact} \frac{\tilde b}{2}\left[1-\sum_{\ell=1}^{N/2} \frac{12 \tilde r_0}{[(2\ell -1)^2\tilde a^2+\tilde b^2]^{5/2}}\right]=0,
\end{equation}
which relates the value of $\tilde b$ to $\tilde a$. Notice that $\tilde b=0$ is always solution of Eq.~\eref{Eq:b:exact}. Values of $\tilde b\neq 0$ are solutions of Eq.~\eref{Eq:b:exact} provided that the linear density $\tilde n$ is larger than some critical value $\tilde n >\tilde n_c$ where
\begin{equation}
\tilde{n}_c=\mathcal E \tilde{r}_0^{-1/5}
\label{eq:nc}
\end{equation}
is the critical density which depends on the trap frequency $\nu_t$, $\mathcal E = (8/(93~\zeta(5)))^{1/5} =0.6078\ldots$ is a constant, and $\zeta$ is the Riemann zeta function. The values of $\tilde b$, computed by solving numerically Eq.~\eref{Eq:b:exact} for different values of $\tilde n$, are shown in Fig.~\ref{fig:b} and compared to values of $\sqrt{\langle y^2\rangle}$ obtained by classical Monte Carlo simulations (see Sec.~\ref{Sec:4} for details on the numerical simulations).

\begin{figure}
a)\\
\includegraphics[width=0.9\columnwidth]{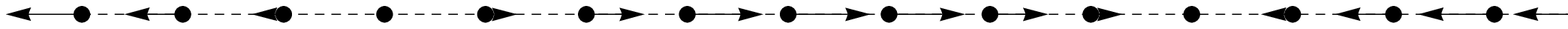}\\
b)\\
\includegraphics[width=0.9\columnwidth]{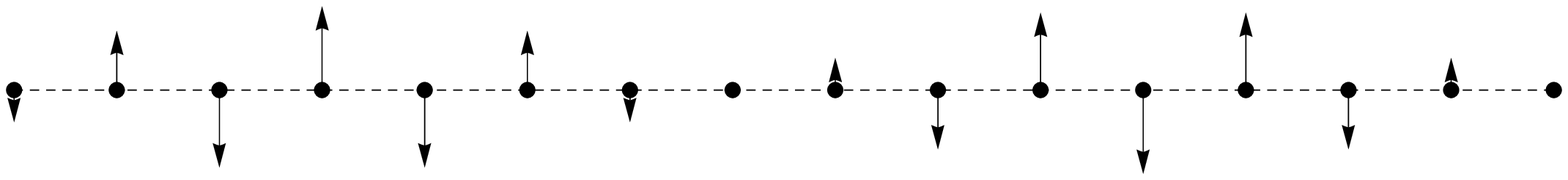}\\
c)\\
\includegraphics[width=0.9\columnwidth]{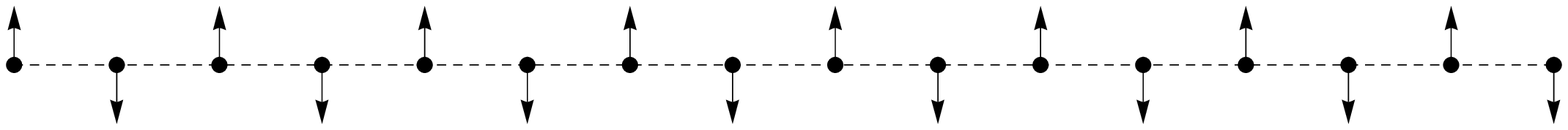}\\
d)\\
\includegraphics[width=0.9\columnwidth]{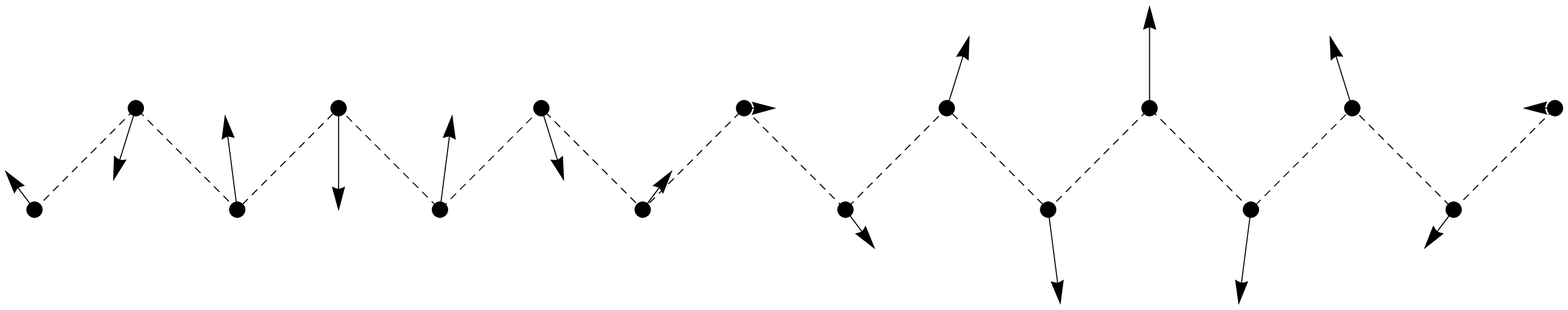}
\caption{Some eigenmodes of the linear chain and of the zigzag structure. a) Linear chain: long-wavelength axial mode at wave vector $k=2\pi/Na$, b) short-wavelength transverse mode at wave vector $k=\pi/a-2\pi/Na$, and (c) transverse mode at $k=\pi/a$ (zigzag mode). d) Zigzag structure: long wavelength eigenmode (this mode can be obtained as a superposition of the modes of the linear chain shown in a) and b)).
} \label{Fig:modes}
\end{figure}

The dispersion relations of the dipolar chain inside a harmonic potential have been determined in Ref.~\cite{Rabl}. In this work we consider a simpler situation, assuming there is no harmonic potential along the $x$ direction, and periodic boundary conditions are imposed at the edges. In this regime the frequencies of the phononic modes of the harmonic crystal are simply found using a phonon-wave {\it ansatz}, see for instance~\cite{Fishman08,Ashcroft}. This ansatz solves exactly the equations of motion for the longitudinal and transverse displacements from the equilibrium positions, $\tilde q_j =\tilde x_j - j\tilde a$ and $\tilde w_j=\tilde y_j$, respectively, which read \begin{eqnarray} \ddot{\tilde q}_j&=&-12 \tilde r_0\sum_{j\neq 0}\frac{1}{j^5\tilde a^5} (\tilde q_j-\tilde q_{j+i}) \label{Eq:linearmotion1}
\\
\ddot{\tilde w}_j&=&-\tilde w_j+3\tilde r_0\sum_{j\neq 0} \frac{1}{j^5\tilde {a}^5}(\tilde w_j-\tilde w_{j+i})
\label{Eq:linearmotion2} \end{eqnarray} One finds two branches, for the axial and transverse excitations, which are given by
\begin{eqnarray}
\omega^2_\parallel(k) &=& \frac{48\tilde r_0}{\tilde a^5} \sum_{j>0}\frac{1}{j^5} \sin^2 \frac{k j a}{2}
\label{Eq:omega_par} \\
\omega^2_\perp(k) &=&1- \frac{12\tilde r_0}{\tilde a^5} \sum_{j>0}\frac{1}{j^5} \sin^2 \frac{k j a}{2}
\label{Eq:omega_perp} \end{eqnarray} where $k=2\pi l/Na$ with $l=0,1,2,\ldots,N-1$. The two branches are analogous to an acoustic ($\omega_{\|}$) and an optical ($\omega_{\perp}$) branch. A few relevant modes are depicted in Fig.~\ref{Fig:modes}.

The acoustic branch, Eq.~(\ref{Eq:omega_par}), is characterized by a sound velocity $c=\sqrt{12\tilde
r_0\zeta(3)/ a^3}$. This value is consistent with the value found from the relation~\cite{Pitaevskii03}
\begin{equation}
mc^2=n \partial \mu /\partial n,
\label{eq:sound}
\end{equation}
where $\mu$ is the chemical potential, $\mu=\partial E/\partial N$ and $E$ is the ground state energy, which at large densities is equal to $E=(N^4/L^3)\zeta(3)C_{dd}/4\pi$ (see \ref{App:A} and Ref.~\cite{Arkhipov05}). From the dispersion relation of the optical branch, Eq.~(\ref{Eq:omega_perp}), one finds the stability condition of the linear chain configuration. In fact, Eq.~(\ref{Eq:omega_perp}) gives imaginary values of the transverse frequency for values of the linear density $\tilde n<\tilde n_c$, with  $\tilde n=1/\tilde a$ and $\tilde n_c$ the critical density. Its value is found by solving the relation ${\rm min}_k(\omega_{\perp})=0$, which gives $\tilde n_c$ in Eq.~\eref{eq:nc} and $k=\pi/a$. The corresponding mode whose frequency becomes zero at $\tilde n_c$ is the zigzag mode (see Fig.~\ref{Fig:modes}c),
\begin{equation}
\label{soft:mode}
\tilde w^{\rm soft}_j=(-1)^j \tilde b/2,
\end{equation}
where $\tilde b$ is the amplitude of the mode when the chain is stable. When one can assume that the dipolar chain exhibits long range order (at very large interaction strengths), the structural instability at $\tilde n=\tilde n_c$ can be characterized as a classical second order phase transition, applying a straightforward extension to dipolar interactions of the analysis presented in Ref.~\cite{Fishman08} for Coulomb interactions. The details are reported in Appendix A.

\section{Monte Carlo methods} \label{Sec:4}

The numerical study of the properties of a many-body system is carried out using Monte Carlo methods. The advantage of these methods is that they permit to evaluate multidimensional integrals over the phase space. In a classical system of dimensionality $D$ this correspond to an integral over $2\times N\times D$ variables ${\bf R} = \{{\bf r}_1, ..., {\bf r}_N;{\bf v}_1,\ldots,{\bf v}_N\}$ with ${\bf r}_j$ and ${\bf v}_j$ being position and velocity of a particle $j$, with $j=1,\ldots,N$. In a quantum system the dimensionality of the phase space is $N\times D$ with ${\bf R} = \{{\bf r}_1, ..., {\bf r}_N\}$ in coordinate representation. The average of some quantity of interest $A({\bf R})$ has the form
\begin{equation}
\langle A\rangle =
\frac{\int ...\int A({\bf R}) p({\bf R})\;d{\bf R}}{\int ...\int p({\bf R})\;d{\bf R}}
\label{A}
\end{equation}
with $p({\bf R})$ the averaging weight. In a classical system at thermal equilibrium position and velocity are independent variables. For a quantity $A({\bf R}) = A({\bf r}_1, ..., {\bf r}_N)$ depending only on particle coordinates, the velocities in Eq.~\eref{A} can be integrated out. As a result the average of interest is performed by only integrating over the position variables, using the averaging weight
\begin{equation}
p_{CLS}({\bf r}_1, ..., {\bf r}_N) = \exp\left\{-\frac{V({\bf r}_1, ..., {\bf r}_N)}{k_BT}\right\}\,
\label{pcls}
\end{equation}
where $V$ is the potential and $T$ the temperature.

In the quantum case the averages are performed using the coordinate representation. In a variational (VMC) calculation the probability distribution is defined by the modulus squared of the trial wave function $\psi_T({\bf r}_1, ..., {\bf r}_N)$,
\begin{equation}
p_{VMC}({\bf r}_1, ..., {\bf r}_N) = |\psi_T({\bf r}_1, ..., {\bf r}_N)|^2.
\label{pvmc}
\end{equation}
The trial wave function $\psi_T$ depends usually on some variational parameters, which can be optimized by minimizing the variational energy of the system. Once the optimal parameters are known, this wave function is used as guiding wave function in Diffusion Monte Carlo (DMC) calculations. The DMC method solves the imaginary-time Schr\"odinger equation for the product of the wave function $\psi$ and the guiding wave function $\psi_T$. For large values of imaginary times the DMC method permits one to sample the probability distribution
\begin{equation}
p_{DMC}({\bf r}_1, ..., {\bf r}_N) = \psi_T({\bf r}_1, ..., {\bf r}_N)\phi_0({\bf r}_1, ..., {\bf r}_N),
\label{pdmc}
\end{equation}
where $\phi_0({\bf r}_1, ..., {\bf r}_N)$ is the ground state wave function (more details on the DMC method can be found, for example, in Ref.~\cite{Boronat94b}).

In both cases, classical (\ref{pcls}) and quantum (\ref{pvmc}), the probability distribution is known analytically and can be written explicitly. This permits one to perform its sampling using the Metropolis algorithm. A trial move converting the probability from $p$ to $p'$ is always accepted if $p'>p$ (in a classical system this means that energy in the new configuration has a lower energy or, being the same, it has a larger value of the wave function in the quantum case), otherwise the move is accepted with probability $p'/p$. We generate a new configuration by displacing one particle by a random distance, which we draw from a Gaussian distribution. The width of the Gaussian distribution is adjusted so that on average we have 50\% acceptance rate. Under those conditions Metropolis algorithm guarantees that the generated set of configurations (Markov chain) has the desired distribution and the average $\langle A\rangle$ is stochastically evaluated as an average over the Markov chain.

We use the Bijl-Jastrow construction of the trial wave function
\begin{equation}
\psi_T({\bf r}_1, ..., {\bf r}_N)=
\prod\limits_{i=1}^N f_1({\bf r}_i)
\prod\limits_{j<k}^N f_2(|{\bf r}_j-{\bf r}_k|).
\label{wf}
\end{equation}
The one-body term is chosen in Gaussian form $f_1({\bf r}) = \exp\{-\alpha y^2/a_{ho}^2\}$. In the limit of quasi-one-dimensional system, where the excitations of the radial harmonic confinement levels are absent, the optimal value $\alpha=1/2$ corresponds to the ground state wave function of a harmonic oscillator. If the system leaves the quasi-one-dimensional regime, the system spreads in the radial direction, and accordingly the value of $\alpha$ is reduced. The two-body term is the same as used in Refs.~\cite{Astrakharchik07a,Astrakharchik08b}
\begin{equation}
f_2(r) =\left\{ \begin{array}{ll}
C_1 K_0(2\sqrt{D a_{ho}/r}), & 0<r<R_{par}\\
C_2\exp\{-C_3 a_{ho}[1/r+1/(L_{x}-r)]\}, & R_{par}\le r < L/2\\
1,& L/2\le r \;, \end{array} \right.
\label{2jastrow}
\end{equation}
with $K_0(r)$ being the modified Bessel function of the second kind. Coefficients $C_1, C_2, C_3$ are fixed by the conditions of the continuity of the wave function and its first derivative. The parameter $0<R_{par}<L/2$ is free and is fixed by a variational procedure (the dependence of the energy on this parameter is rather weak and in many cases the choice $R_{par}=L/4$ is sufficient). When the distance between two particles is small, the influence of other particles can be neglected. As a result, $f_2({\bf r})$ is well approximated by the solution of the two-body scattering problem (see short distance behavior of (\ref{2jastrow})). This makes the calculation to be stable even for a divergent interaction potential. We symmetrize the argument of the long-range part as $1/r + 1/(L_x-r)$. In this way we ensure the zero derivative of the trial wave function at the boundary $r=L_x/2$ according to the periodic boundary conditions.

DMC averaging (\ref{pdmc}) provides exact estimators for operators commuting with the Hamiltonian. In particular, the ground state energy is found in a statistically exact way. Instead, density profiles and other functions of particle positions are obtained as ``mixed'' estimators $\langle A\rangle_{DMC} = \langle \psi_T|A|\phi_0\rangle/\langle \psi_T|\phi_0\rangle$ which are biased by a particular choice of the trial wave function. Ideally, one has to find a pure estimator, {\it i.e.} averages over the ground state $\langle A\rangle = \langle \phi_0|A|\phi_0\rangle/\langle \phi_0|\phi_0\rangle$. The optimized trial wave function is expected to be close to the true ground state wave function, so that the difference $|\delta\psi\rangle = |\psi_T\rangle-|\phi_0\rangle$ is small and one can use the mixed estimator together with the variational one $\langle A\rangle_{VMC} = \langle \psi_T|A|\psi_T\rangle/\langle\psi_T|\psi_T\rangle$ (obtained according to (\ref{pvmc})) to extrapolate the pure estimator. The two extrapolation procedures
\begin{eqnarray}
\label{extrapolation1}
\langle A\rangle &\approx& 2\langle A\rangle_{DMC} - \langle A\rangle_{VMC}\\
\langle A\rangle &\approx& \langle A\rangle_{DMC}^2/\langle A\rangle_{VMC}
\label{extrapolation2}
\end{eqnarray}
have the same order of accuracy and contains errors which are quadratic in $|\delta\psi\rangle$. Alternatively, for local quantities one can use more involved techniques of pure estimators \cite{Sarsa02}. For simplicity we will use the rules (\ref{extrapolation1}-\ref{extrapolation2}) and check that both estimators agree within the accuracy of interest.

\section{Thermal fluctuations around the classical equilibrium positions}
\label{Sec:5}

The structures discussed in Sec.~\ref{Sec:3} were determined by simply evaluating the classical equilibrium positions of the potential energy, Eq.~\eref{eq:tildeV}, as a function of the linear density and of the transverse trapping frequency. In this case, we identified a sharp condition,  $\tilde{n}_c=\mathcal E \tilde{r}_0^{-1/5}$, as given by Eq.~(\ref{eq:nc}), such that for $\tilde n<\tilde n_c$ the equilibrium positions are along a line (linear chain), otherwise are arranged along multiple lines. For an interval $\tilde n_c<\tilde n<\tilde n_M$, with $\tilde n_M$ a fixed value, the structure is a zigzag chain. Thermal fluctuations will smear the sharp transition at $\tilde n_c$. In this section we compute their effect around $\tilde n_c$ for various values of the temperature using a classical Monte Carlo calculation.

\begin{figure}
\begin{center}
\includegraphics[width=0.5\columnwidth]{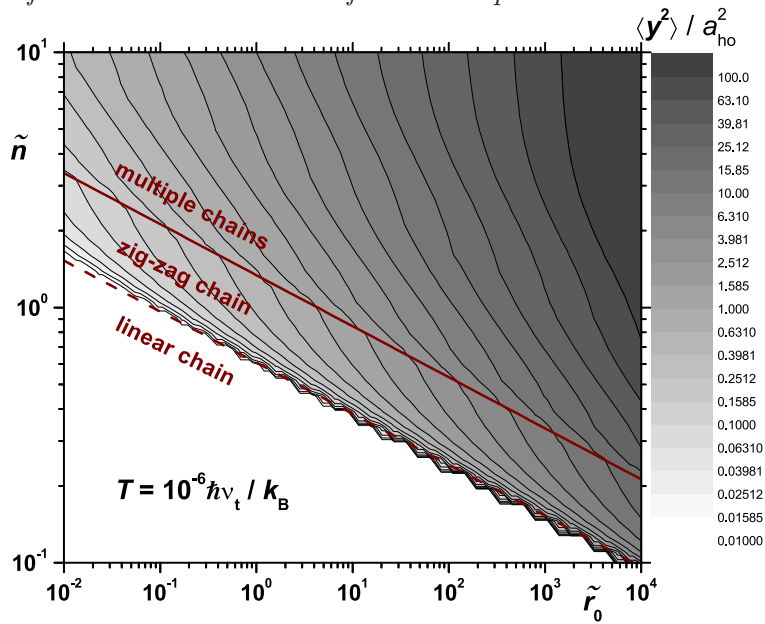}
\caption{(color online) Contour plot of the mean transverse size of the system $\langle y^2\rangle$ at $T = 10^{-6}\hbar\nu_t/k_B$ as a function of the parameters $\tilde{r}_0$ and $\tilde{n}$. The contour lines correspond to fixed values of $\langle y^2\rangle$ (in units of $a_{ho}^2$), the scale is reported on the right side of the plot. The dashed line, obtained from solving Eq.~(\ref{eq:nc}), separates the linear from the zigzag structures; the solid line separates the zigzag from the multiple chain structures. The calculations are made for a system of $N=20$ particles, where we covered the plane ($\tilde n$ - $\tilde r_0$) with a grid of $50\times 50$ size, each point corresponding to a run with a different choice of parameters. Effects of the finite grid are visible in Monte Carlo data in the vicinity of the dashed line.
} \label{Fig:PhT1e-6}
\end{center}
\end{figure}

\begin{figure}
\begin{center}
\includegraphics[width=0.5\columnwidth]{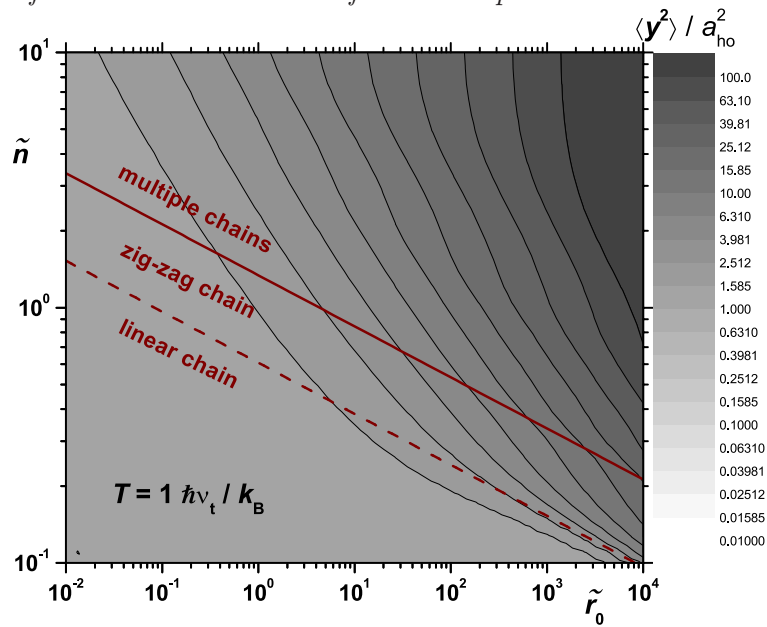}
\caption{(color online) Same as Fig.~\ref{Fig:PhT1e-6} but for $T = \hbar\nu_t/k_B $.
} \label{Fig:PhT1}
\end{center}
\end{figure}

For later convenience, we report the temperature in units of the energy $E_{ho}$, which we set $E_{ho}=\hbar\nu_t$.  In order to study the transition linear -- zigzag structure we evaluate the transverse variance $\langle y^2\rangle$ and report the numerical results as a function of the linear density $\tilde n$ and the interaction strength $\tilde{r}_0$ in a contour plot for two different values of the temperature, see Figs.~\ref{Fig:PhT1e-6}-\ref{Fig:PhT1}. In addition, in the figures we plot two lines corresponding to the transition from a linear to a zigzag (dashed line), and from a zigzag to a multiple chain (solid line), as obtained by finding the classical equilibrium positions of the potential~\eref{eq:tildeV}, and plotted for comparison.

Figure~\ref{Fig:PhT1e-6} reports the result obtained setting $T = 10^{-6}\hbar\nu_t/k_B$. Here, one can still identify a sharp transition from linear to zigzag, and from zigzag to multiple chains, according to the predictions given by Eq.~\ref{eq:nc}. Such situations could be found in setups with extremely high transverse frequencies, although it is questionable whether such trap could be experimentally realized. Figure~\ref{Fig:PhT1} displays the case $T=\hbar\nu_t/k_B$, showing that at this temperature the transition is completely smeared out.

We now study how thermal effects modify the pair correlation function, which is related to the probability of finding two particles separated by a distance $(x,y)$, and is here defined as
\begin{equation}
g_2(x,y)=\langle {\mathcal N}({\bm \rho}_1)\;  \mathcal N({\bm \rho}_2)\rangle\,,
\end{equation}
where $\mathcal N({\bm \rho})=\sum_i^N \delta({\bm\rho}-{\bm\rho}_i)$ is the two-dimensional density, and $g_2(x,y)$ depends only on the relative distances $x=x_1-x_2$ and $y=y_1-y_2$. Here, integrating out the radial coordinate from the two-dimensional density gives the linear density $\int\mathcal N({\bm \rho})\;{\rm d}y = n(x)$ and its average is equal to the linear density $\langle n(x)\rangle=n$.

\begin{figure}
\begin{center}
\includegraphics[width=0.49\columnwidth]{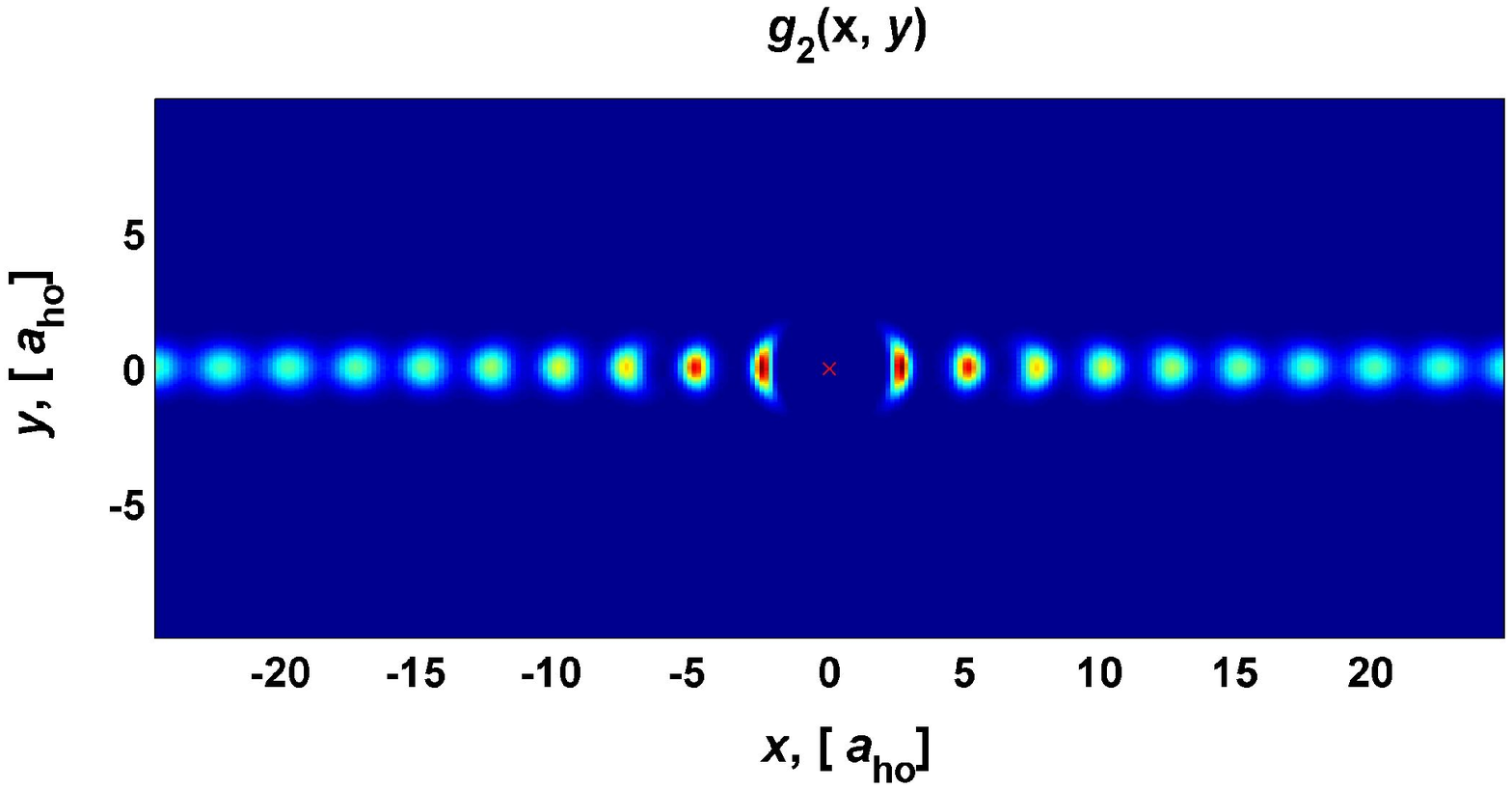}
\includegraphics[width=0.40\columnwidth]{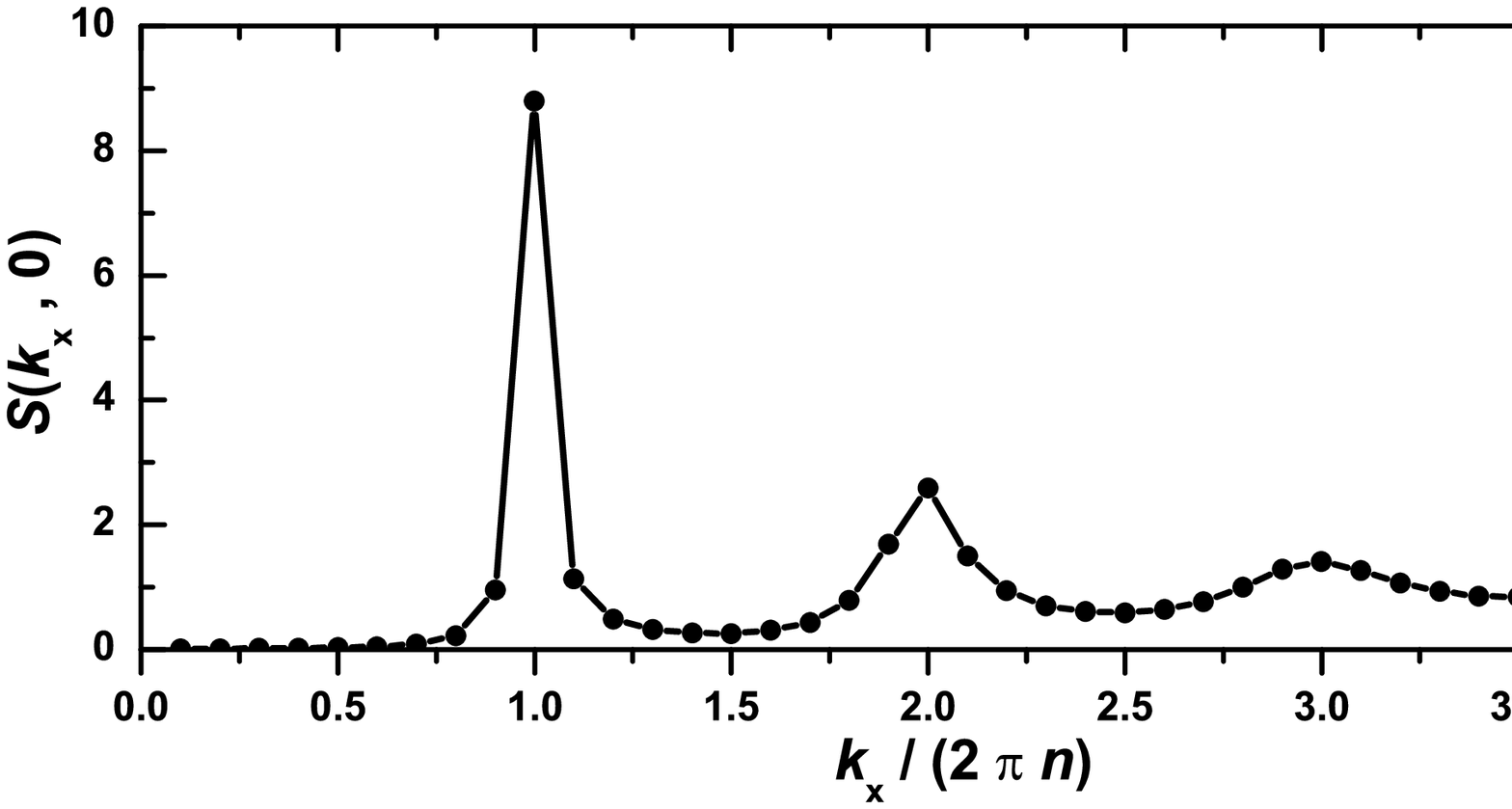}
\includegraphics[width=0.49\columnwidth]{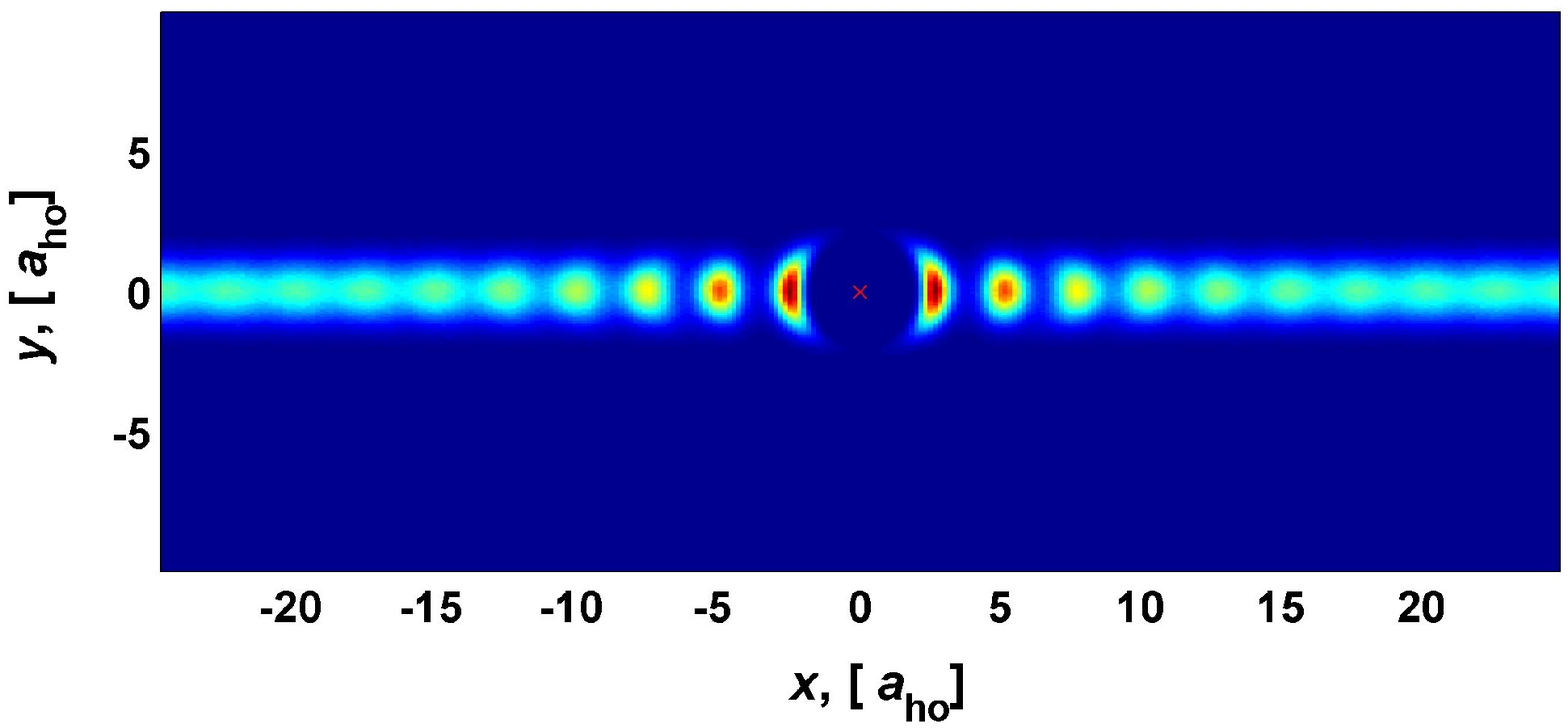}
\includegraphics[width=0.40\columnwidth]{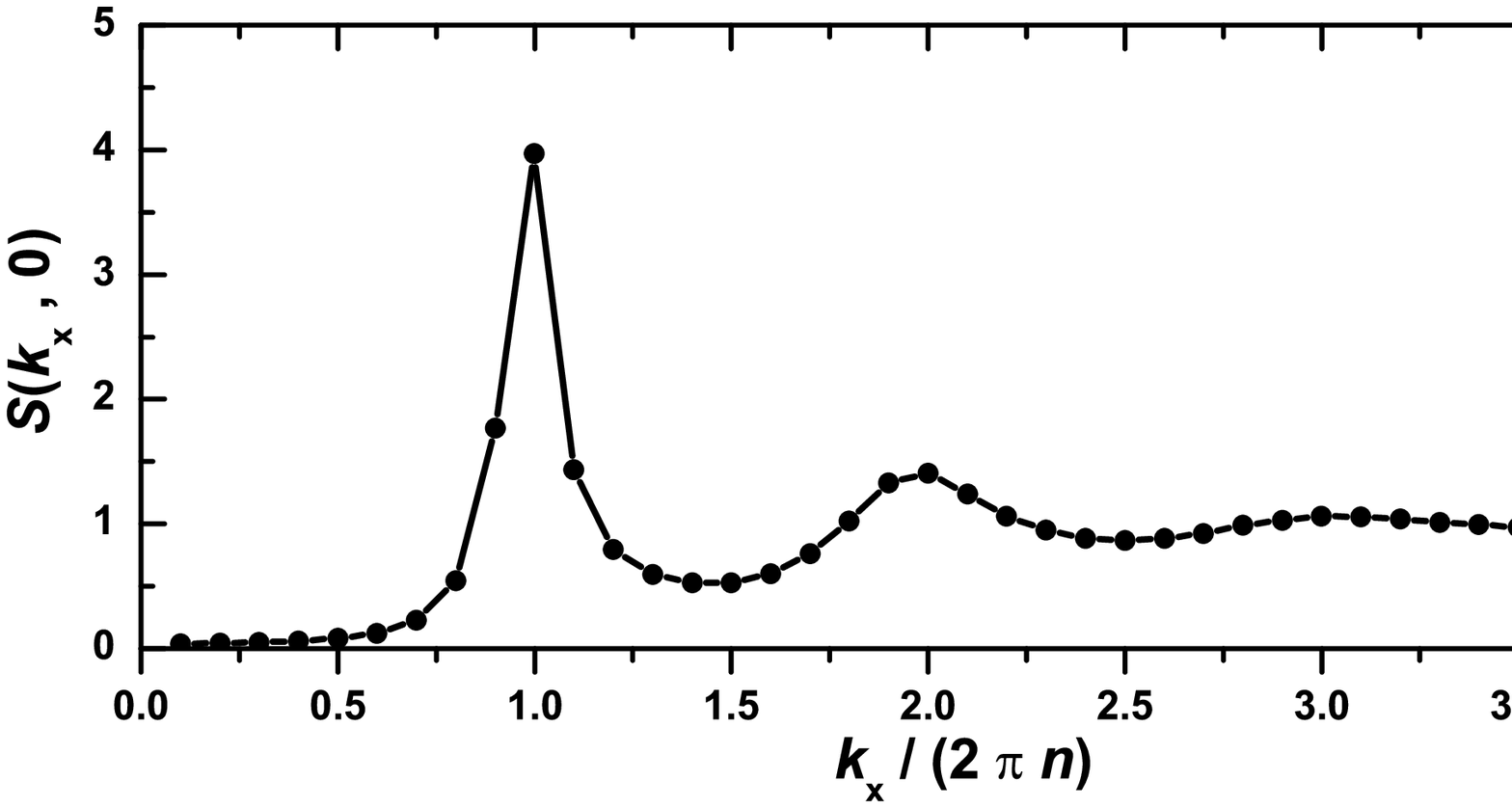}
\includegraphics[width=0.49\columnwidth]{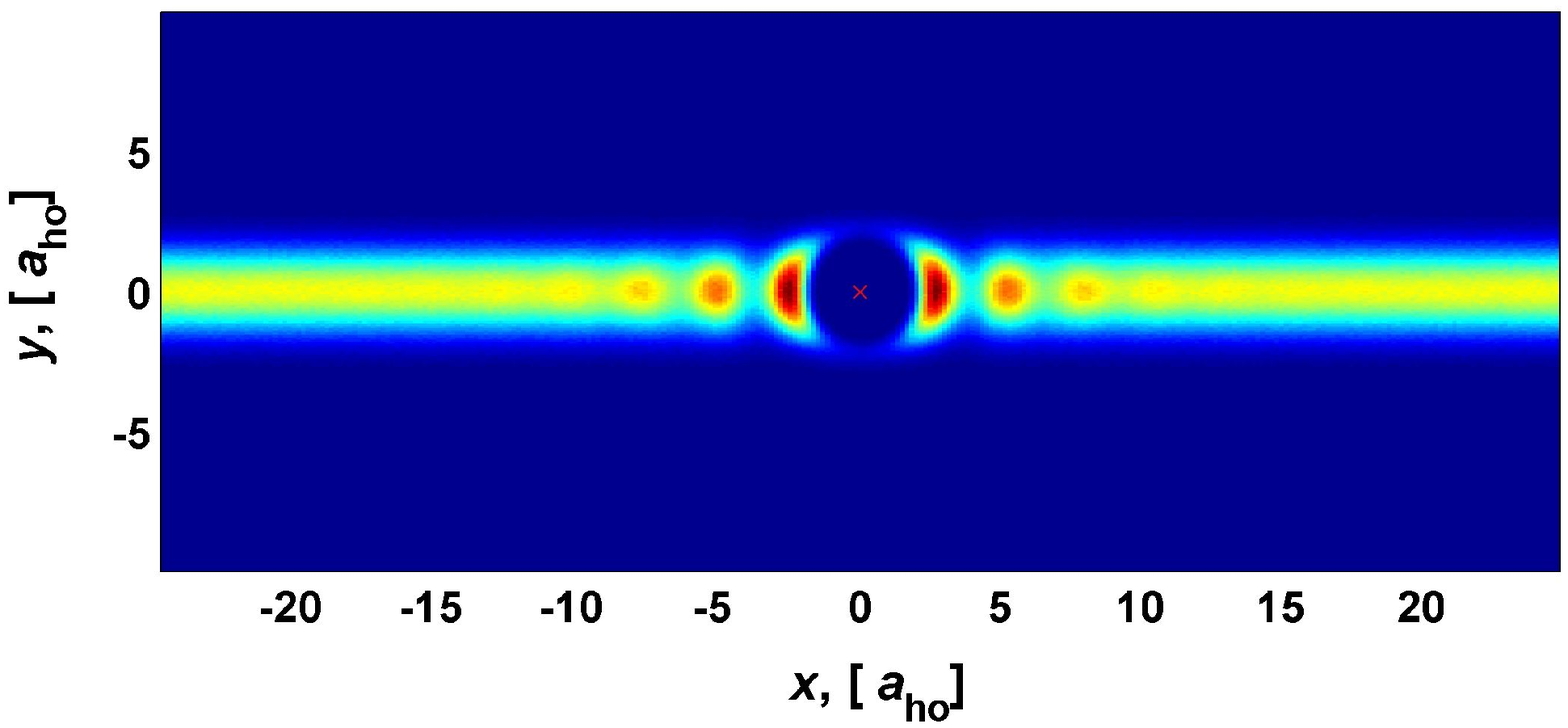}
\includegraphics[width=0.40\columnwidth]{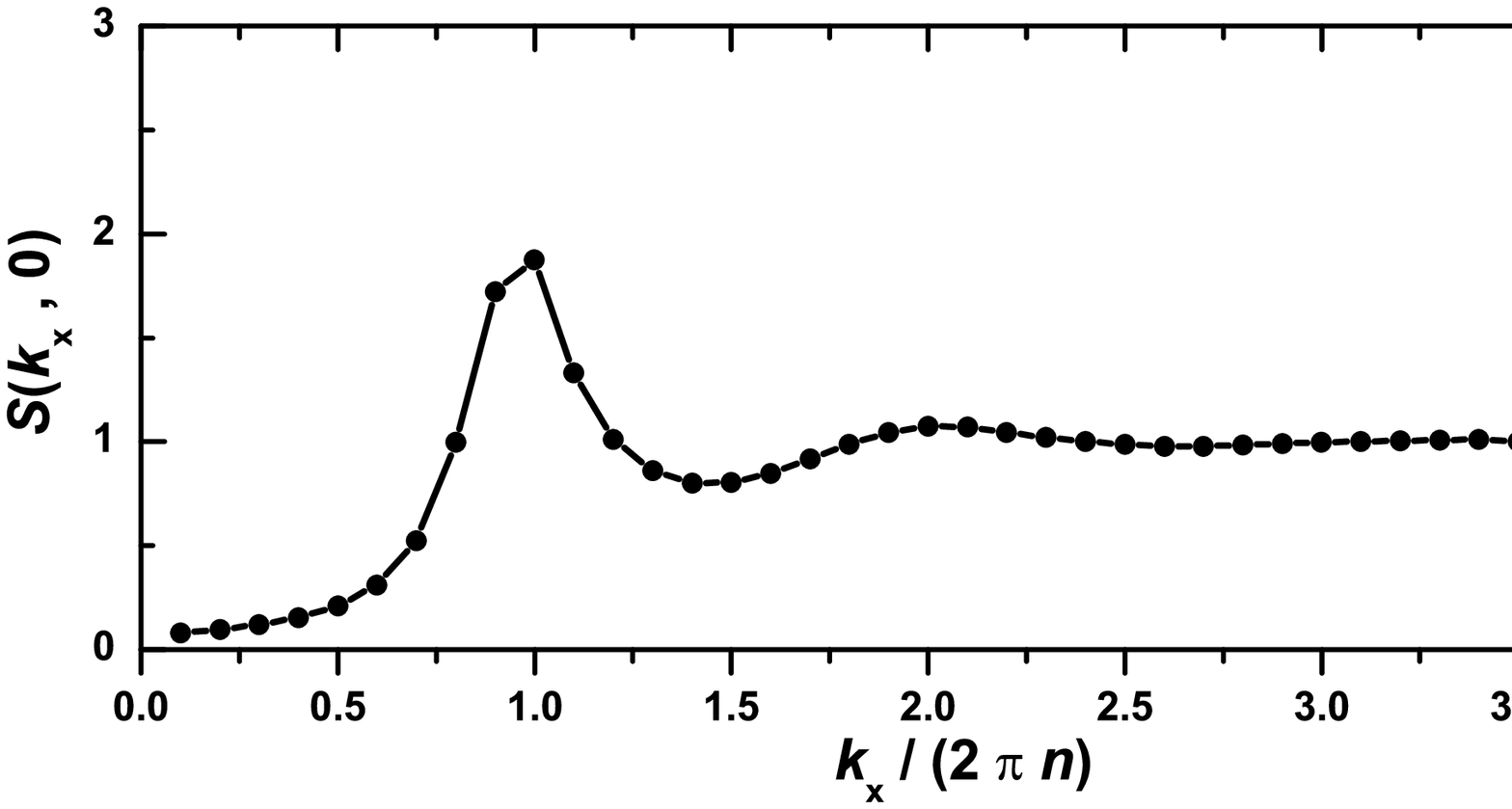}
\includegraphics[width=0.49\columnwidth]{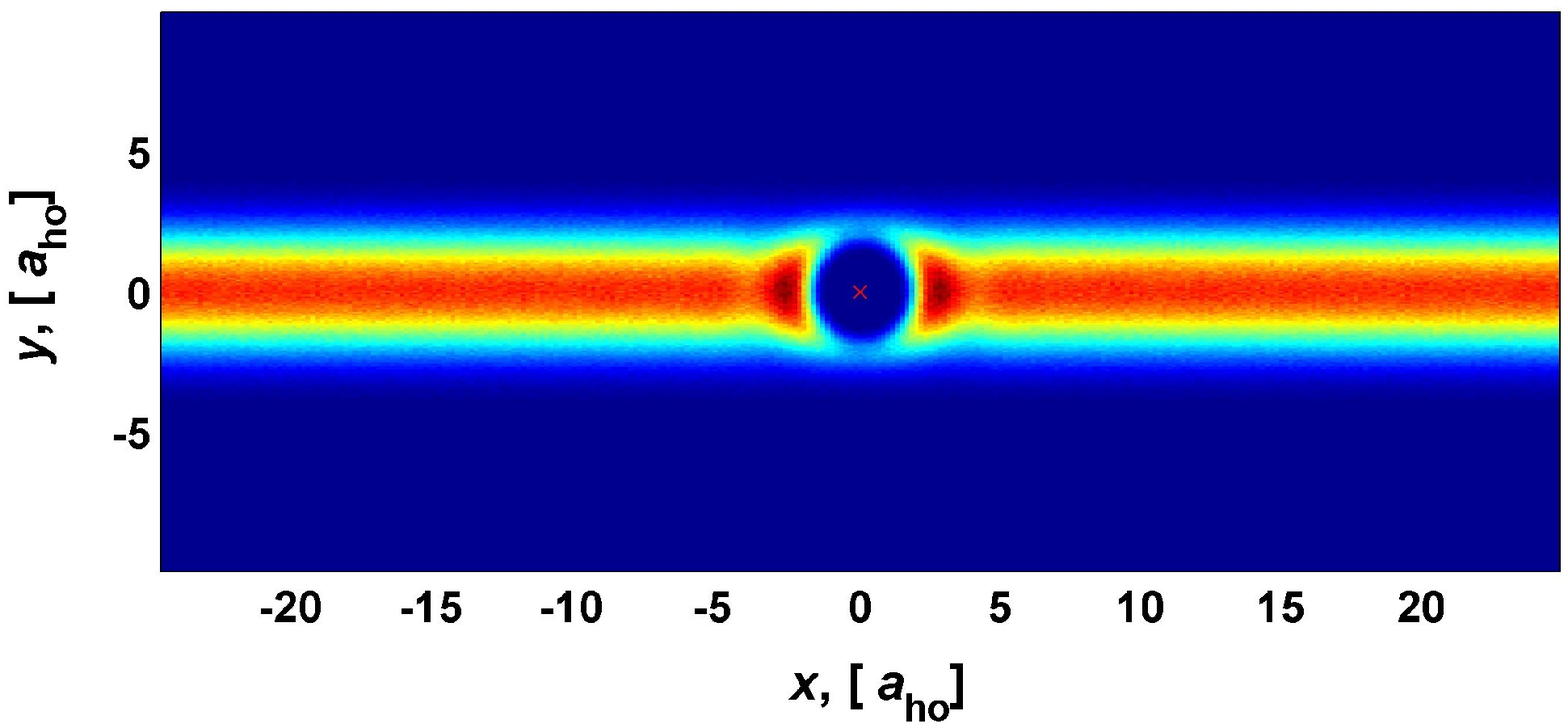}
\includegraphics[width=0.40\columnwidth]{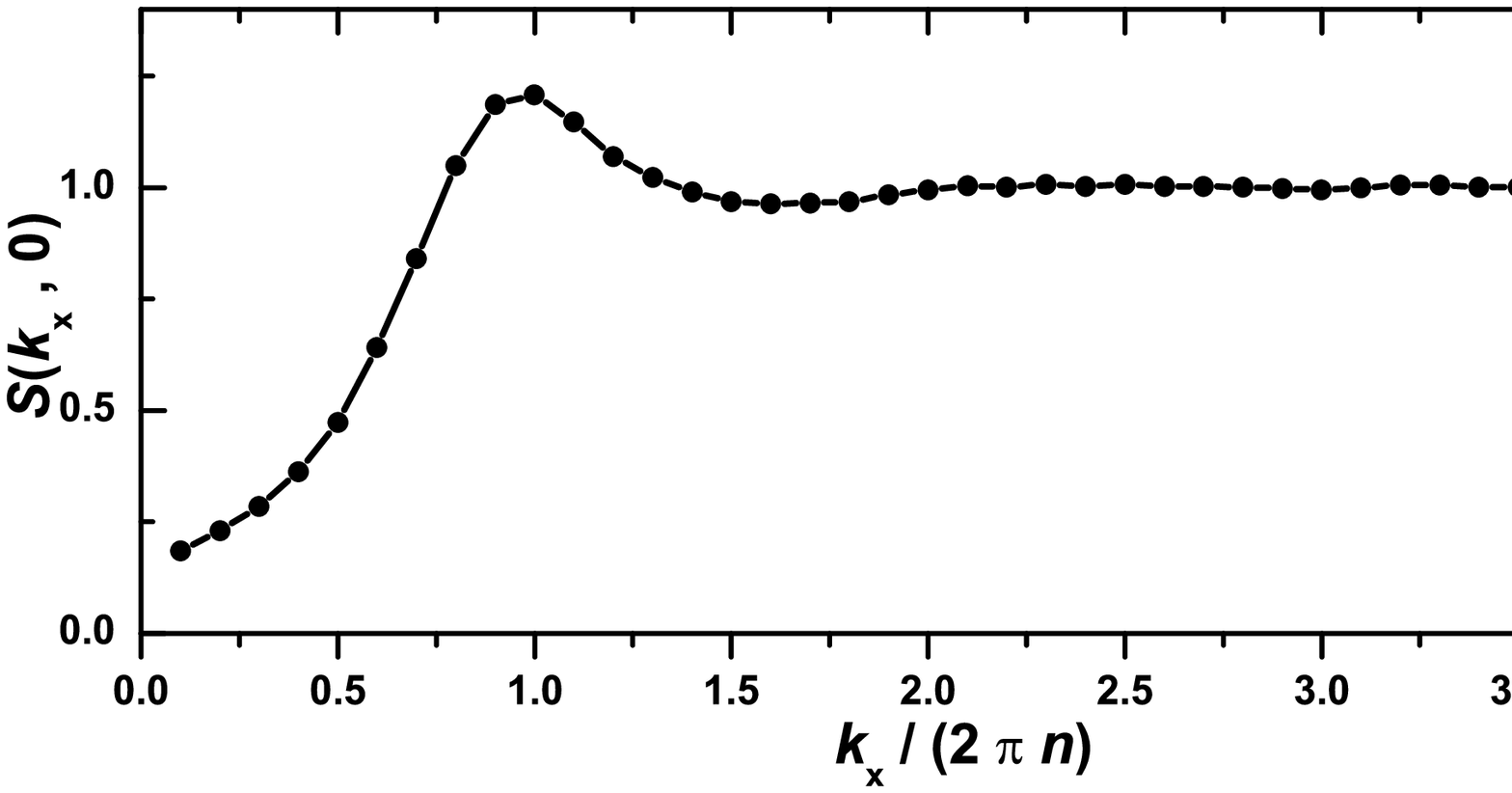}
\caption{(color online) Pair distribution function $\langle \tilde n (\bm \varrho)\tilde n (0)\rangle$ (left) and static structure factor ($S_k$) for the linear structures (right). Temperatures are (rows from upper to low) $T=0.05$, $T=0.1$, $T=0.2$ and $T=0.5 \hbar\nu_t/k_B$. Number of particles is $N=20$, $\tilde r_0 = 7.5$, $\tilde n=0.4$}
\label{fig:PDlinear}
\end{center}
\end{figure}

The pair correlation function is displayed in Fig.~\ref{fig:PDlinear} for various values of the temperature. Here, the ``hole'' for $x=0$ and $y=0$ is due to the hard-core repulsive interaction potential, which forbids two particles being at the same position. There is a clear strong correlation between neighboring particles, which is lost as the temperature and the interparticle distance increase. This is a manifestation of the absence of a true crystalline order.

Information over the properties of the system can be inferred by means of Bragg spectroscopy, giving access to the static structure factor. The latter is given by
\begin{equation}
S({\bf k}) = \frac{1}{N}\langle {\mathcal N}_{\bf k}{\mathcal N}_{\bf -k} \rangle
\label{eq:Sk}
\end{equation}
where ${\mathcal N}_{\bf k} = \int e^{i{\bf k\bm\rho}} {\mathcal N}({\bm\rho})d{\bm\rho} = \sum_{j=1}^N \exp\{i{\bf k\bm\rho}_j\}$ is the Fourier transform of the two-dimensional density ${\mathcal N}({\bm\rho})$. The right column of plots in Fig.~\ref{fig:PDlinear} displays the static structure factor, corresponding to the pair correlation function plotted on the left side. Here, one sees the appearance of a high peak, at $k_x=2\pi/a$, and of a second peak, at $k_x=4\pi/a$, at low temperatures. The peaks would correspond to Bragg scattering by an ordered structure at interparticle spacing $a$, and are thus signatures of very strong correlations in the system. Their height diminishes, while they become broader, as temperature increases and the axial structure is lost. The disappearance of the second peak at higher temperatures is typical of a gas phase.

\begin{figure}
\begin{center}
\includegraphics[width=0.49\columnwidth]{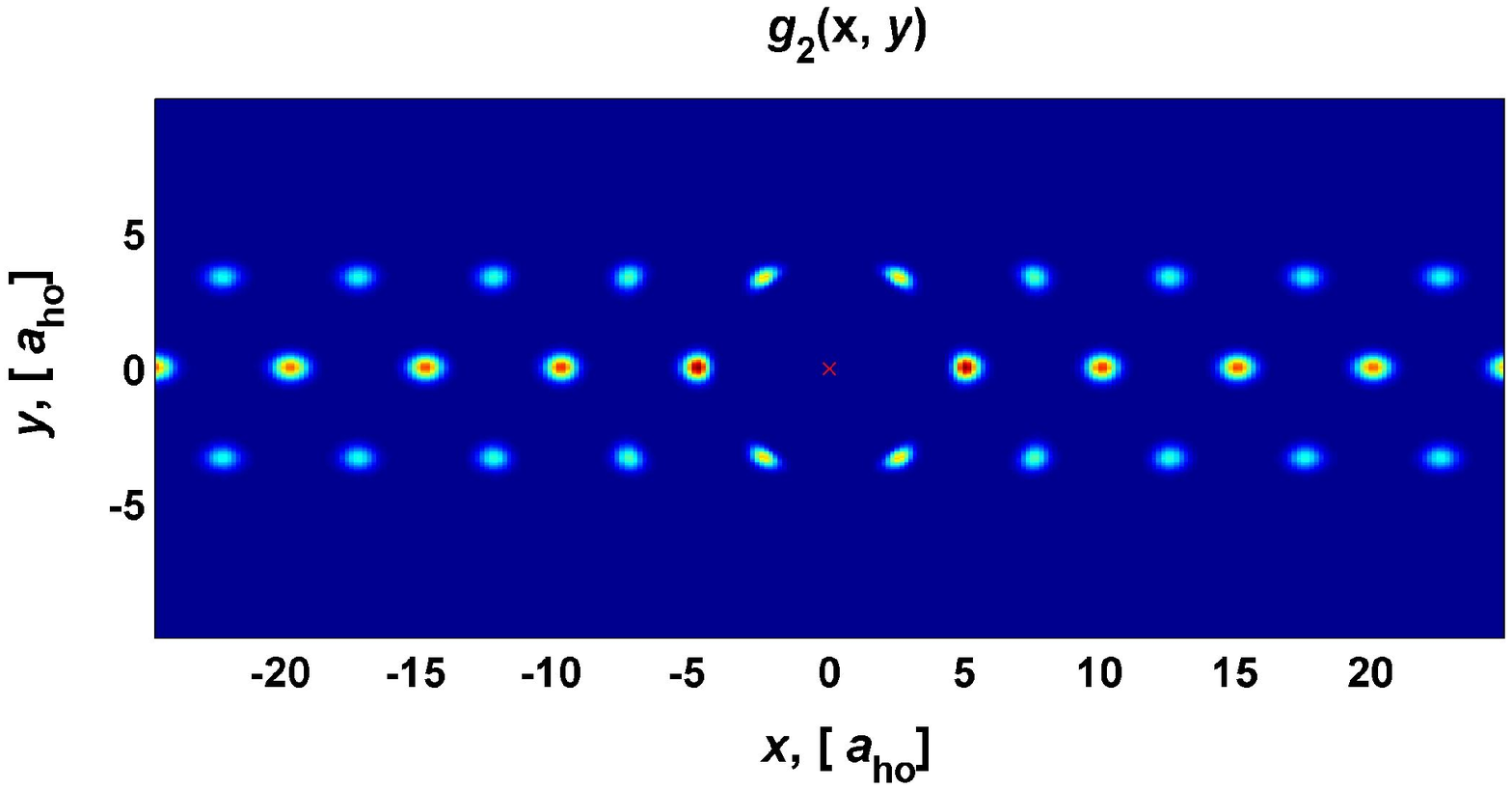}
\includegraphics[width=0.49\columnwidth]{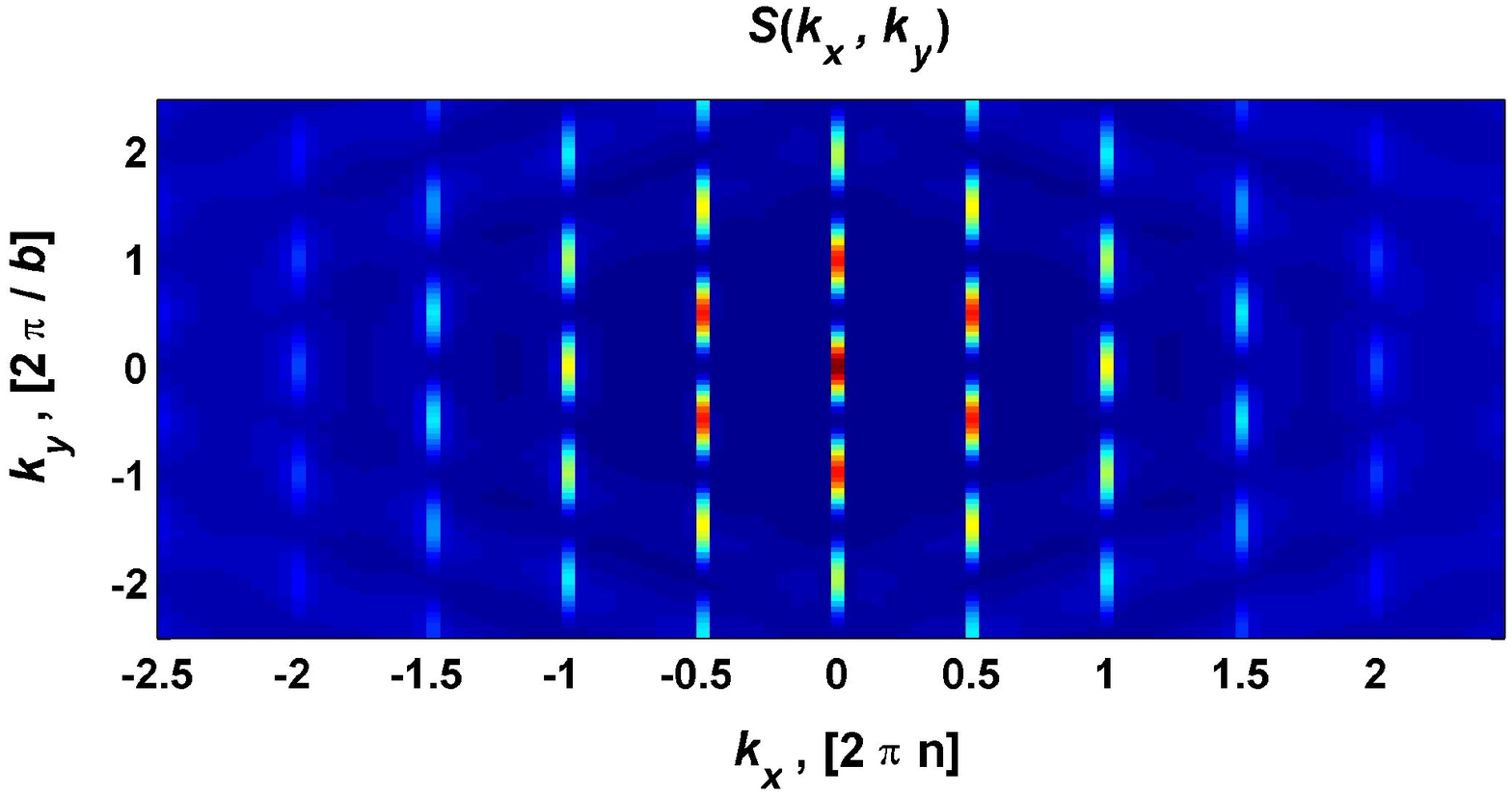}
\includegraphics[width=0.49\columnwidth]{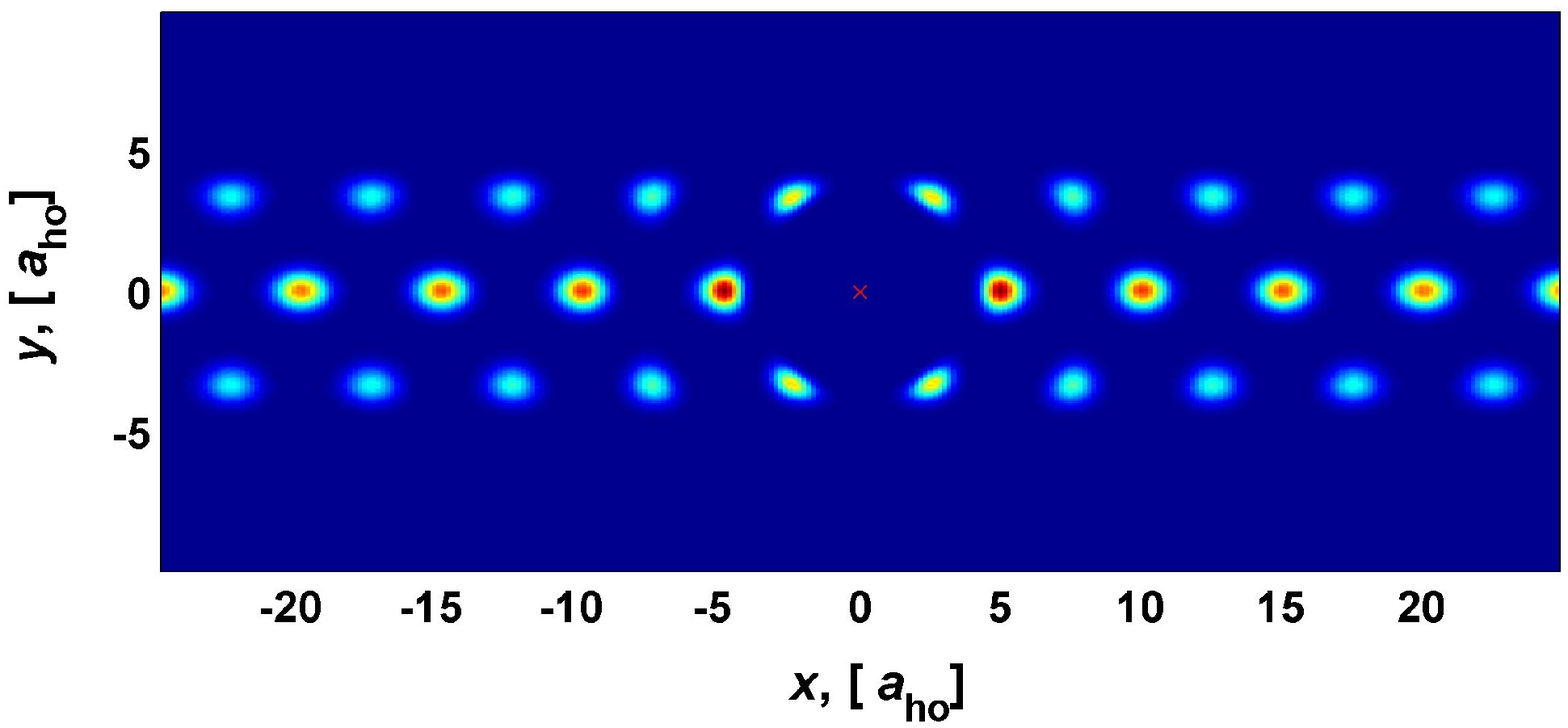}
\includegraphics[width=0.49\columnwidth]{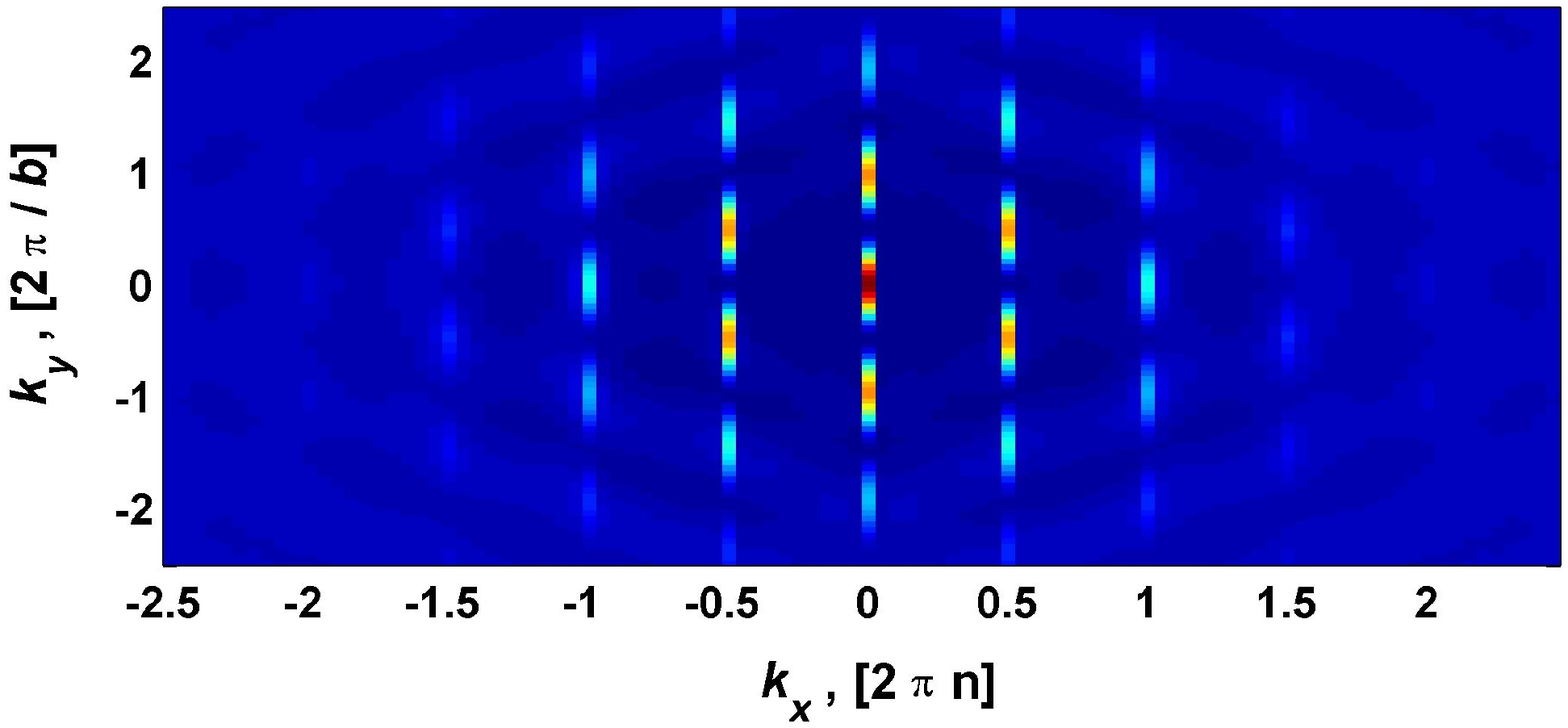}
\includegraphics[width=0.49\columnwidth]{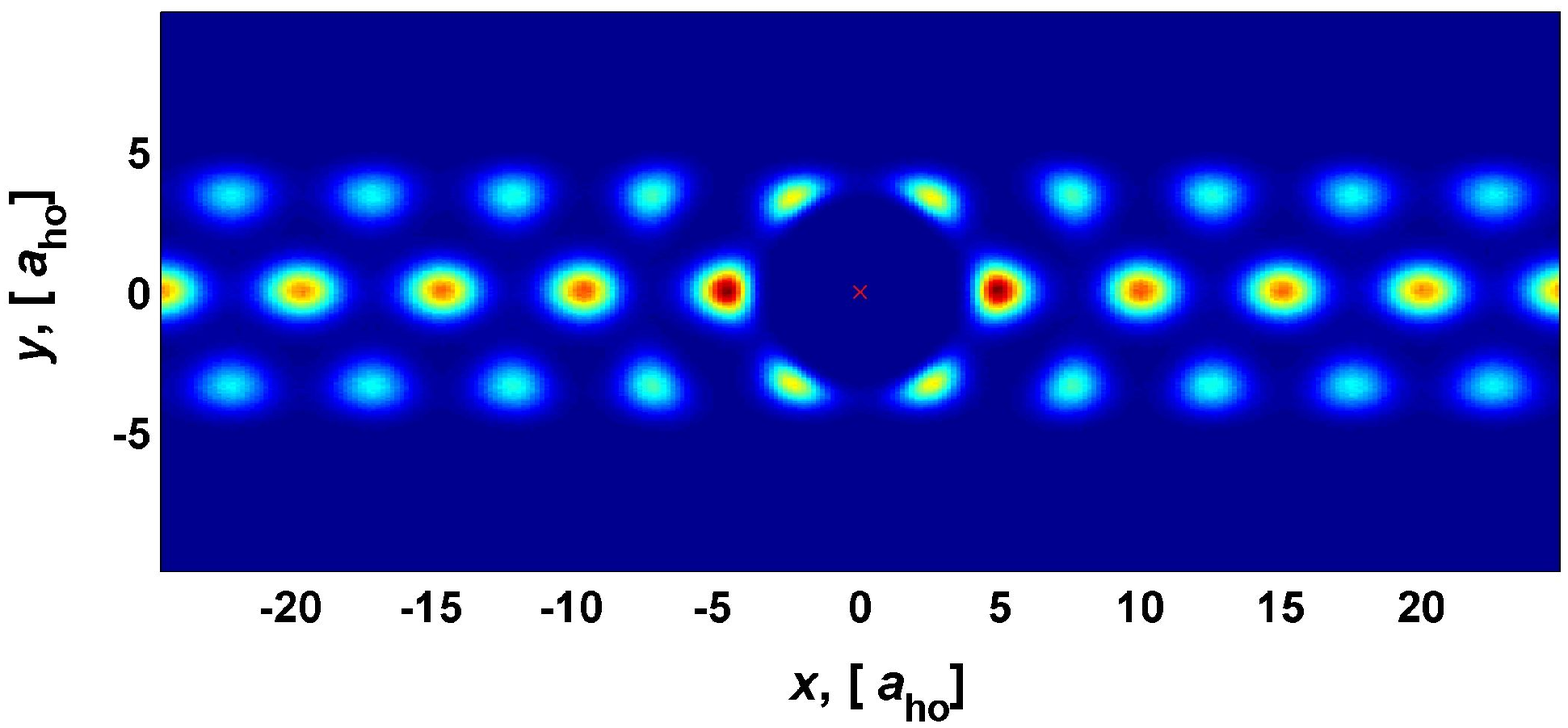}
\includegraphics[width=0.49\columnwidth]{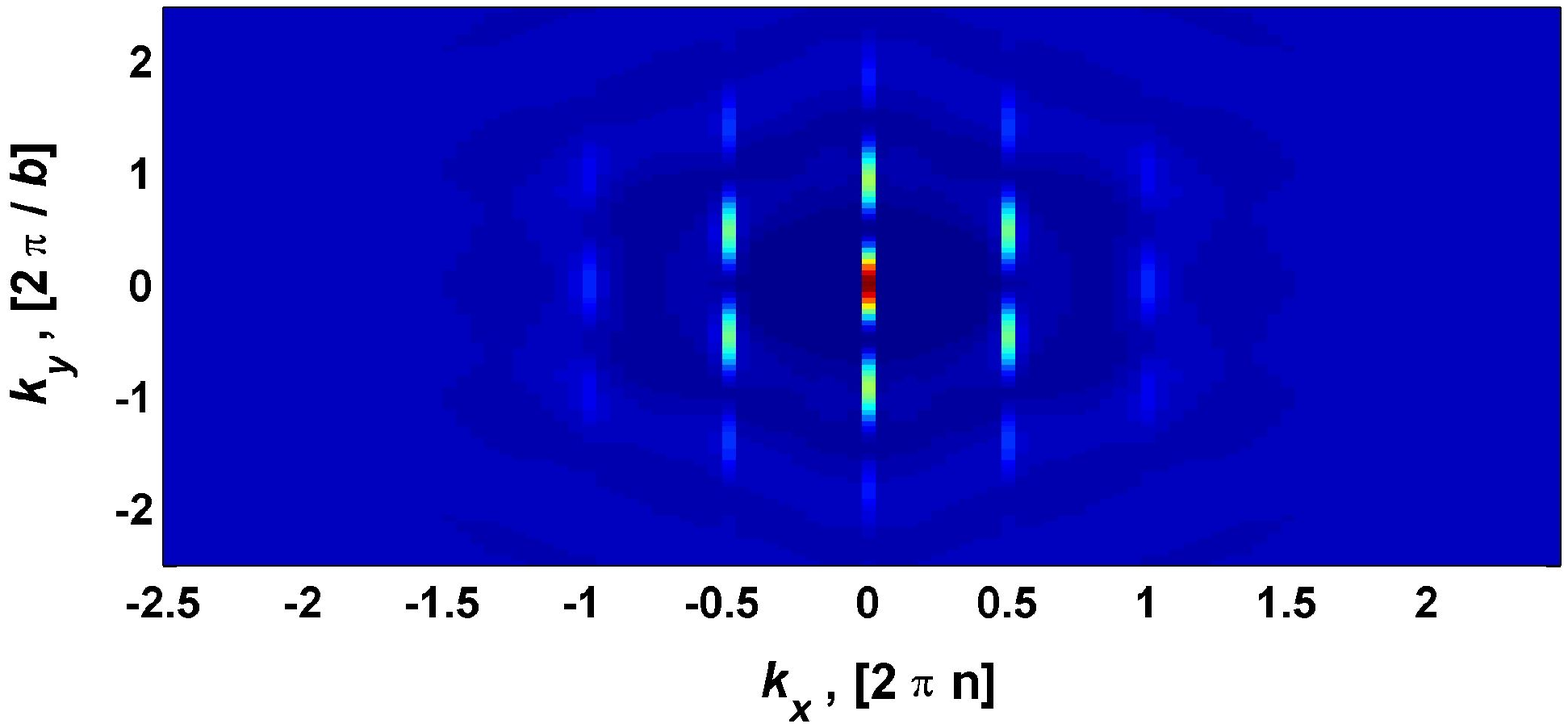}
\includegraphics[width=0.49\columnwidth]{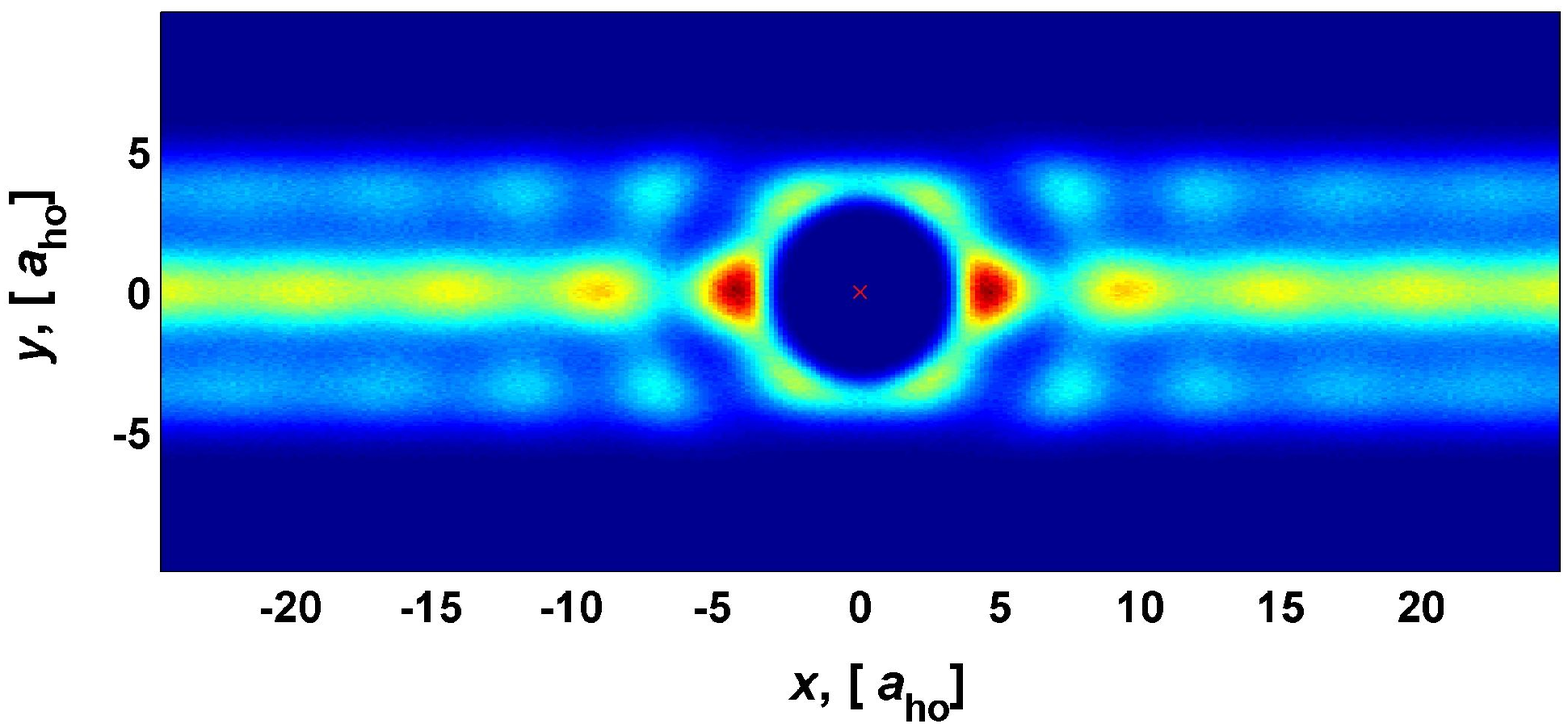}
\includegraphics[width=0.49\columnwidth]{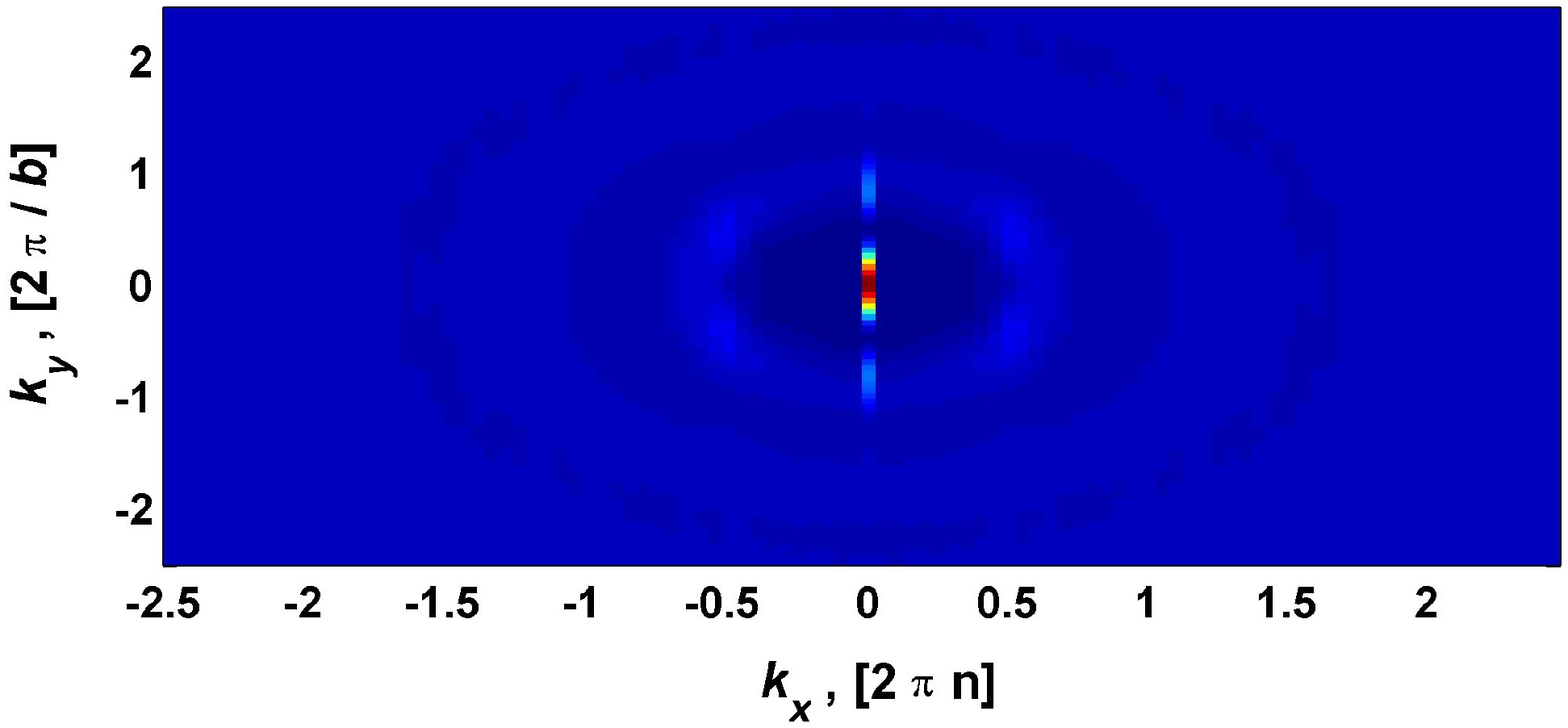}
\caption{(color online) Same as in Fig.~\ref{fig:PDlinear}, but with $\tilde r_0 = 100$ and $\tilde n=0.4$, warranting the zigzag regime. Temperatures are (rows from upper to low) $T=0.05$, $T=0.1$, $T=0.2$ and $T=0.5 \hbar\nu_t/k_B$. Differing from Fig.~\ref{fig:PDlinear}, the static form factor is now a mesh plot as a function of $k_x$ (in units of $2\pi/a$) and of $k_y$ (in units of $2\pi/b$).}
\label{fig:PDzigzag}
\end{center}
\end{figure}

Figure~\ref{fig:PDzigzag} displays the pair correlation function and the structure form factor in the parameter regime, in which the zigzag structure is expected at ultralow temperatures. As before, the plots of $g_2(x,y)$ show a hole for the zero separation, caused by the dipolar repulsion. Moreover, one finds a crystal-like ordering at very low temperatures, $T\approx 0.05 \hbar\nu_t/k_B$, which disappears as the temperature is increased by a factor of 10. In addition, at low temperatures one observes large probability to find two particles with the same transversal displacement, such that $y_i-y_j=0$, with doubled periodicity with respect to the linear chain. One also observes additional peaks corresponding to correlations in the $y$ direction, corresponding to the doubled chain (zigzag) structure. As in one half of the cases the particle of reference belongs to the upper (lower) chain, there are peaks in $g_2(x,y)$ for $y = - 2b$ ($y=2b$). The peaks at $y = 0$ have larger amplitudes as they are sum of the contributions from both upper and lower chains.

The static structure factor is plotted as a function of the two-dimensional wave vector $(k_x,k_y)$. We note that different boundary conditions in $x$ and $y$ lead to different accessible values of momentum.\footnote{Here, periodic boundary conditions along $x$ lead to a discretization in momentum, with unit step equal to $2\pi/L = (2\pi n) / N$, while in $y$ direction the momentum is continuum. In practice this means that typically the width of the crystal-like peaks along $k_x$ cannot be determined, while along $k_y$ the peaks have well-defined width. In the thermodynamic limit the discretization disappears and the momentum is continuous in all directions.} At low temperatures the static structure factor reminds the geometry of a two-dimensional crystal, and naturally shows how a one-dimensional structure transforms into a two-dimensional one. There is a number of directions in which well-defined peaks are observed. The peaks in direction $(k_x,k_y) = (\pi/a, 0)$ are similar to the ones shown in Fig.~\ref{fig:PDlinear} for the linear chain. The peaks in the transverse direction $(k_x,k_y) = (0,\pi/b)$ or in diagonal directions $(k_x,k_y) = (\pi/a,\pi/b)$, {\it etc.} are signatures of the realization of a zigzag structure. As the temperature is increased, the number of peaks is decreased until only the central peak is left. We note, however, that the presence of the central peak is somehow trivial as it counts the number of particles in the system.

\begin{figure}
\begin{center}
\includegraphics[width=0.5\columnwidth, angle=-90]{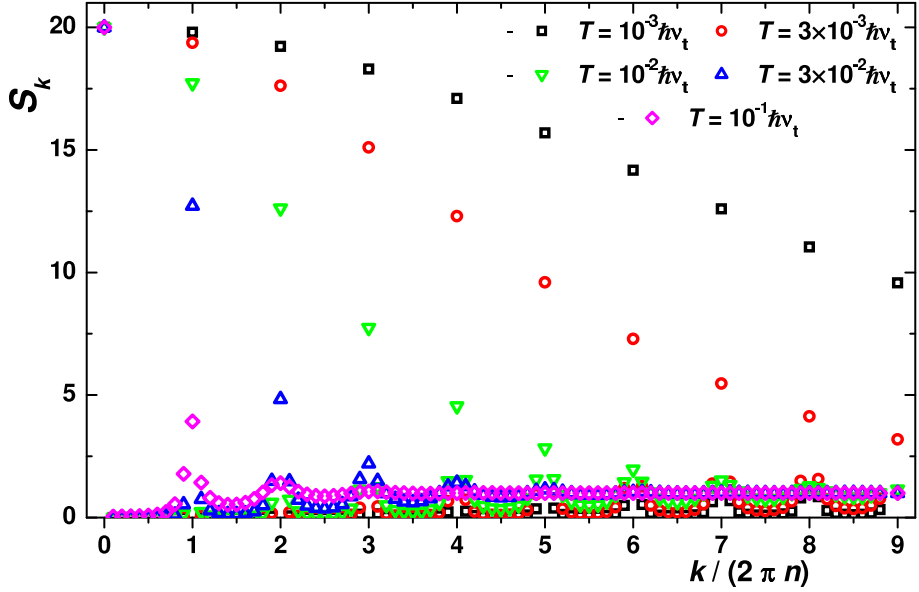}
\caption{(color online) Static structure factor of a linear chain of polar molecules. Parameters are the same is in Fig.~\ref{fig:PDlinear}, while the temperatures are $T  = 10^{-4}; 3\times 10^{-4};  10^{-3}; 3\times 10^{-3}; 10^{-2}~\hbar\nu_t/k_B$ (see plot).}
\label{fig:SkPeak}
\end{center}
\end{figure}

Finally, we study how the ordering in the linear chain is destroyed as the temperature is increased from $T=10^{-3}\hbar\nu_t/k_B$, up to $T=0.1\hbar\nu_t/k_B$, by looking at the features of the structure form factor. In Fig.~\ref{fig:SkPeak} $S_k$ is plotted as a function of $k_x$ for $N=20$ particles, for various values of the temperature. The peak at $k=0$ counts the number of particles and its height equals to 20 in all cases. The height of the other peaks, at multiples of $2\pi/a$, diminishes as $T$ is increased. In particular, it decreases with the number of the peak. Moreover, the height of the peaks seems to follow a power-law dependence as a function of the peak number. These observations will be compared with the effect of quantum fluctuations on ordering at $T=0$.

\section{The quantum ground state \label{Sec:6}}

\begin{figure} \begin{center}
\includegraphics[width=0.5\columnwidth, angle=-90]{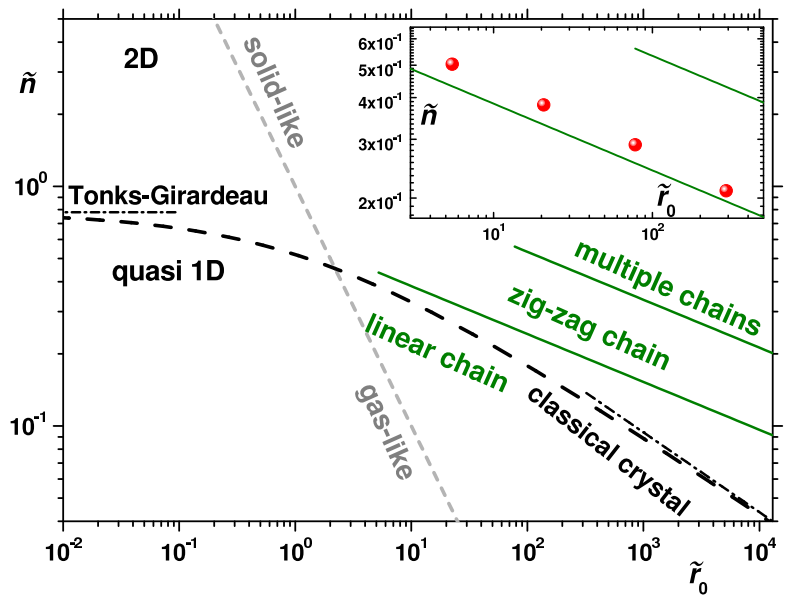}
\caption{(color online) Ground--state phase diagram as a function of the parameters $\tilde{r}_0$ and $\tilde{n}$. The dashed black line corresponds to the curve $E^{1D}/N=\hbar\nu_t$ and identifies the two regimes, where the dynamics is essentially one- or two-dimensional. The Tonks--Girardeau and the classical crystal limits are explicitly indicated in the plot (dashed-dotted lines). The short-dashed grey line, at $\tilde{n}\tilde{r}_0=1$, separates gas- and solid-like phases. These lines are to be intended as indicators, the transition of the gas from one phase to another being a crossover. The green solid lines separate different quasi-crystalline structures in the classical system. Inset: portion of the phase diagram in the solid-like regime and close to the region, where the classical transition linear-zigzag structure is found (solid line). The red circles are results of quantum Monte Carlo calculations and correspond to the appearance of a double-peak structure in the radial density profile (see details in Ref.~\cite{Astrakharchik08b}). The size of the symbols denotes the error bars.}
\label{Fig:1} \end{center} \end{figure}

The definition of a one-dimensional ground state density distribution of dipoles in the quantum regime requires the existence of an energy gap for exciting the transverse motion, such that at $T=0$ the relevant excitations of the dipolar gas are essentially along the longitudinal axis. In order to quantify this statement we first introduce the characteristic lengths of the system. The confinement in the transverse direction sets the characteristic length (oscillator length)
\begin{equation}
a_{ho}=\sqrt{\hbar\;/\;m\nu_t},
\end{equation} 
which determines the width of the single-particle ground-state wave packet, and which we choose as unit length. We extend the classical model of Eq.~\eref{eq:hamtilde} by taking the unit energy $E_{ho} =\hbar\nu_t$ in which now the classical variables ${\bm \varrho}$ and ${\bm \pi}$ are replaced by the corresponding quantum conjugate operators. The potential energy $\tilde V$ is given by the Eq.~\eref{eq:tildeV}.
For a one-dimensional configuration, when the transverse excitations are frozen out, the parameter $r_0$ gives the characteristic length scale of interactions, and has to be compared to the length scale of quantum fluctuations set by $n^{-1}$, where $n$ is the linear density. The system is in a gas or quasi-crystal phase depending on whether the product $n r_0$ is much smaller or much larger than unity. For $nr_0\gg 1$ the dipolar gas is in the quasi-ordered phase with ground-state energy $E_{\rm cr}^{(1D)}= N(nr_0)^3\zeta(3)\hbar^2/(mr_0^2)$, which is the potential energy of a classical crystal~\cite{Arkhipov05} (see also \ref{App:A}). In the quantum gas regime, for $nr_0\ll 1$, the properties of the system are well described by the Tonks--Girardeau model~\cite{Girardeau60} with ground state energy $E_{\rm TG}^{(1D)}=N\pi^2\hbar^2n^2/(6m)$ (the expression for $E_{\rm TG}^{(1D)}$ does not depend on the strength of the dipolar interaction, as in this regime the interparticle distance is large and the potential interaction enters only through the positions of the nodes of the ground state wave function, while the potential energy can be neglected). These two regimes are shown in the phase diagram in Fig.~\ref{Fig:1}, displaying the various phases of the ground state of the dipolar system as a function of $\tilde{r}_0$ and $\tilde{n}=na_{ho}$, and are separated by the grey short-dashed line, which indicates the curve $nr_0=\tilde n\tilde r_0=1$.

Dynamically the system is one-dimensional when the chemical potential $\mu$ is much smaller than the level spacing of the transverse oscillator, $\mu \ll \hbar\nu_t$. In this regime, the quantum state of the system can be described within the Luttinger liquid formalism, as it was shown in Ref.~\cite{Citro07}, and no true long-range order is found at finite densities~\cite{Citro08}. The curve $E^{(1D)}/N=\hbar\nu_t$ separates the one- and two-dimensional phases and is shown in Fig.~\ref{Fig:1} by the black dashed line. In the quasi-ordered phase, for $nr_0\gg 1$, this corresponds to the inequality $E_{\rm cr}^{(1D)}/N\ll \hbar\nu_t$, and which leads to the relation $n\ll r_0^{-1/3}$. In the quantum gas regime, for $nr_0\ll 1$, the condition of being one-dimensional is $E_{\rm TG}^{(1D)}/N\ll \hbar\nu_t$, which is equivalent to the requirement $na_{ho}\ll 1$.

The phase diagram in Fig.~\ref{Fig:1} can be experimentally explored by changing the density $\tilde n$, moving vertically, and by varying the transverse frequency $\nu_t$, moving parallel to the straight short-dashed line. Typical parameters of experiments with one--dimensional gases are given for instance in Ref.~\cite{Kinoshita05}: for $a_{ho} \approx 35$ nm, $60$ --- $400$ atoms per tube of length varying between $15$ and $50\mu$m, one finds $\tilde r_0 \approx 0.07$, $\tilde n \approx 0.04$ --- $0.3$. Among all atom species with which condensation has been reached, $^{52}$Cr has the largest value of $r_0 = 2.4$ nm~\cite{Pfau08}. Larger values of $r_0$ can be reached using polar molecules such as CO, ND$_3$, HCN, CsCl with $r_0 = 5$~nm --- 340~$\mu$m, permitting one to cover regions of the phase diagram up to the classical region ($\tilde n\tilde r_0\gg 1$).

\section{Effects of quantum fluctuations on ordering}\label{Sec:7}

In this section we present numerical results for the pair correlation function and the structure form factor of the quantum gas of bosons at $T=0$, focusing on the solid-like regime and studying in particular the parameter regimes, where classically the linear and zigzag chains are observed. Now, the structural properties are defined from the interplay between the quantum kinetic term, dipolar interactions and the confinement potential in the Hamiltonian~(\ref{eq:hamtilde}). The results are compared with the classical case at finite temperature, discussed in Sec.~\ref{Sec:5}.

\begin{figure}
\begin{center}
\includegraphics[width=0.49\columnwidth]{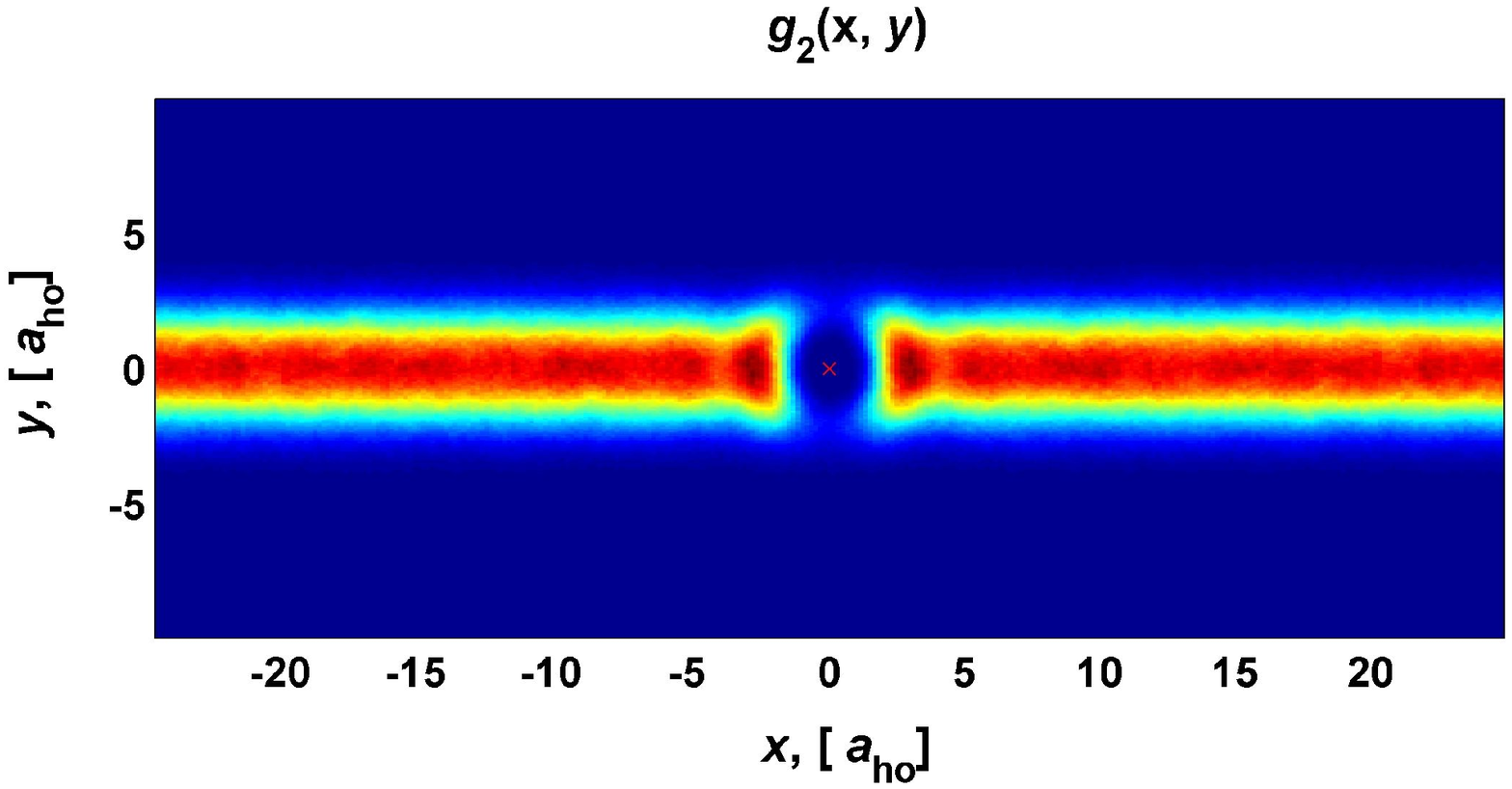}
\includegraphics[width=0.4\columnwidth]{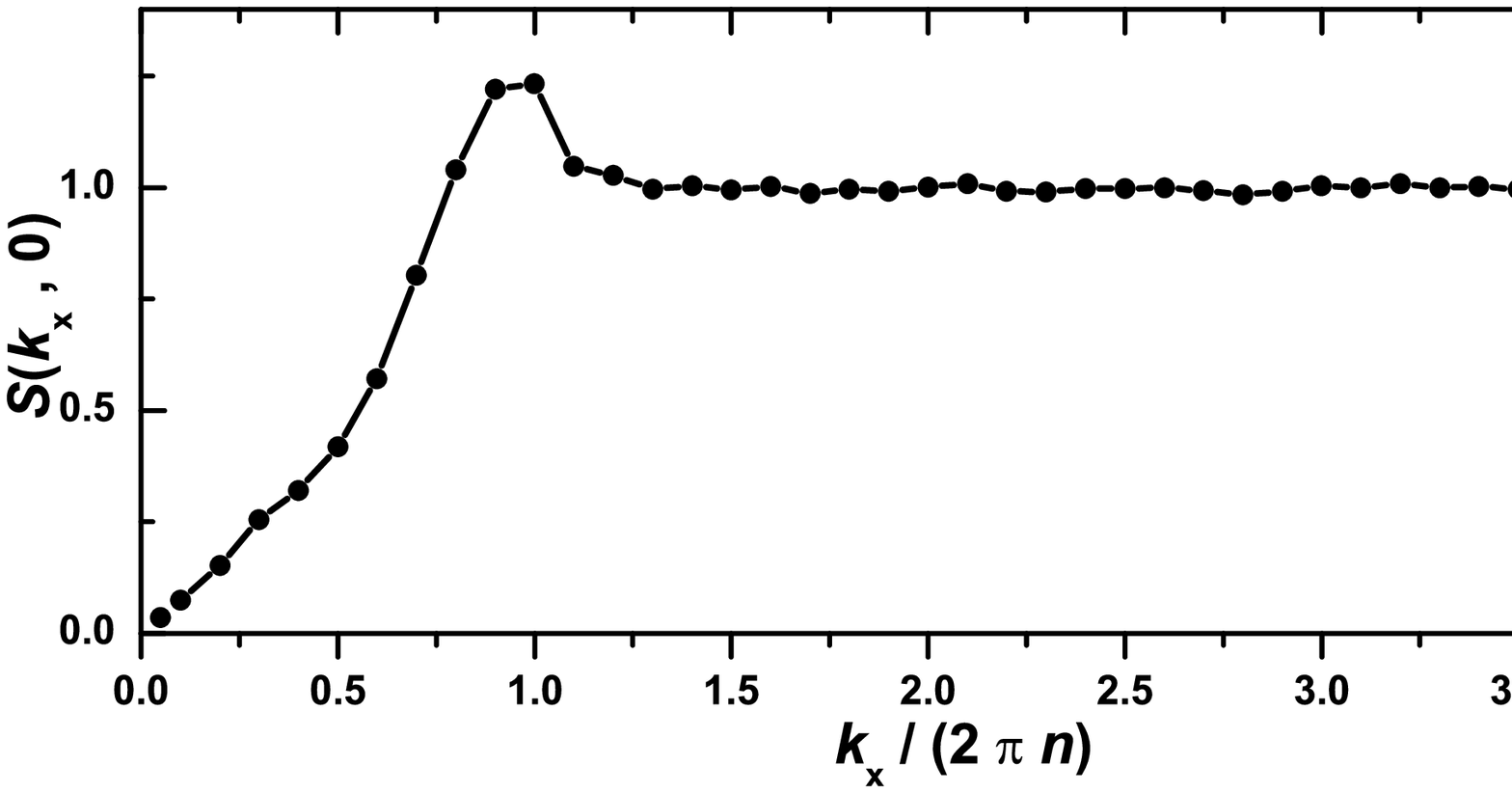}
\caption{(color online) Pair distribution function $g_2(x,y)$ (left) and static structure form factor $S_k$ for the quantum gas of dipolar bosons at $T=0$, with $N=20$ particles, $\tilde n=0.4$ and $\tilde r_0=7.5$. The results are obtained with a Quantum Monte Carlo simulation, see Sec.~\protect\ref{Sec:4}. Extrapolation (\ref{extrapolation1}) is used to remove bias due to a particular choice of trial wave function}
\label{fig:QNT}
\end{center}
\end{figure}

Figure~\ref{fig:QNT} displays the pair correlation function and the static structure factor of the quantum system at zero temperature. For the same parameter regimes in a classical system, we observed at very low temperature quasi-crystalline linear structures, see Fig.~\ref{fig:PDlinear}. The results in Fig.~\ref{fig:QNT} hence clearly show that quantum fluctuations significantly modify the phase of the system. No features of quasi-ordering can be observed: the pair correlation function is almost uniform (apart for the hole at $x=y=0$~\footnote[1]{As discussed in Sec.~\ref{Sec:5}, the hole comes from the short-range divergence of the repulsive potential. This can be easily understood by considering the lower limit of the integral giving the potential energy of dipolar interactions: $\int V(r) g_2({\bf r})\; d{\bf r} \propto \int g_2({\bf r})/|r|^3\; d{\bf r}$. This energy diverges if $g_2({\bf r})$ does not vanish at short distances. Consequently the pair correlation function has to vanish when two particles meet both in classical and quantum system. In a quantum system this also means that the two-body Jastrow term has to vanish at short distances, which is in agreement with the behavior of the two-body scattering solution used in the construction of the trial wave function (\ref{2jastrow}).}), and the static structure form factor exhibits a small peak at the wavevector, corresponding to the mean linear density $n$, while no further peaks at multiples of $k=2\pi n$ are visible. An analogous situation is observed in the classical system at sufficiently high temperatures, compare with Fig.~\ref{fig:PDlinear}.

\begin{figure}
\begin{center}
\includegraphics[width=0.49\columnwidth]{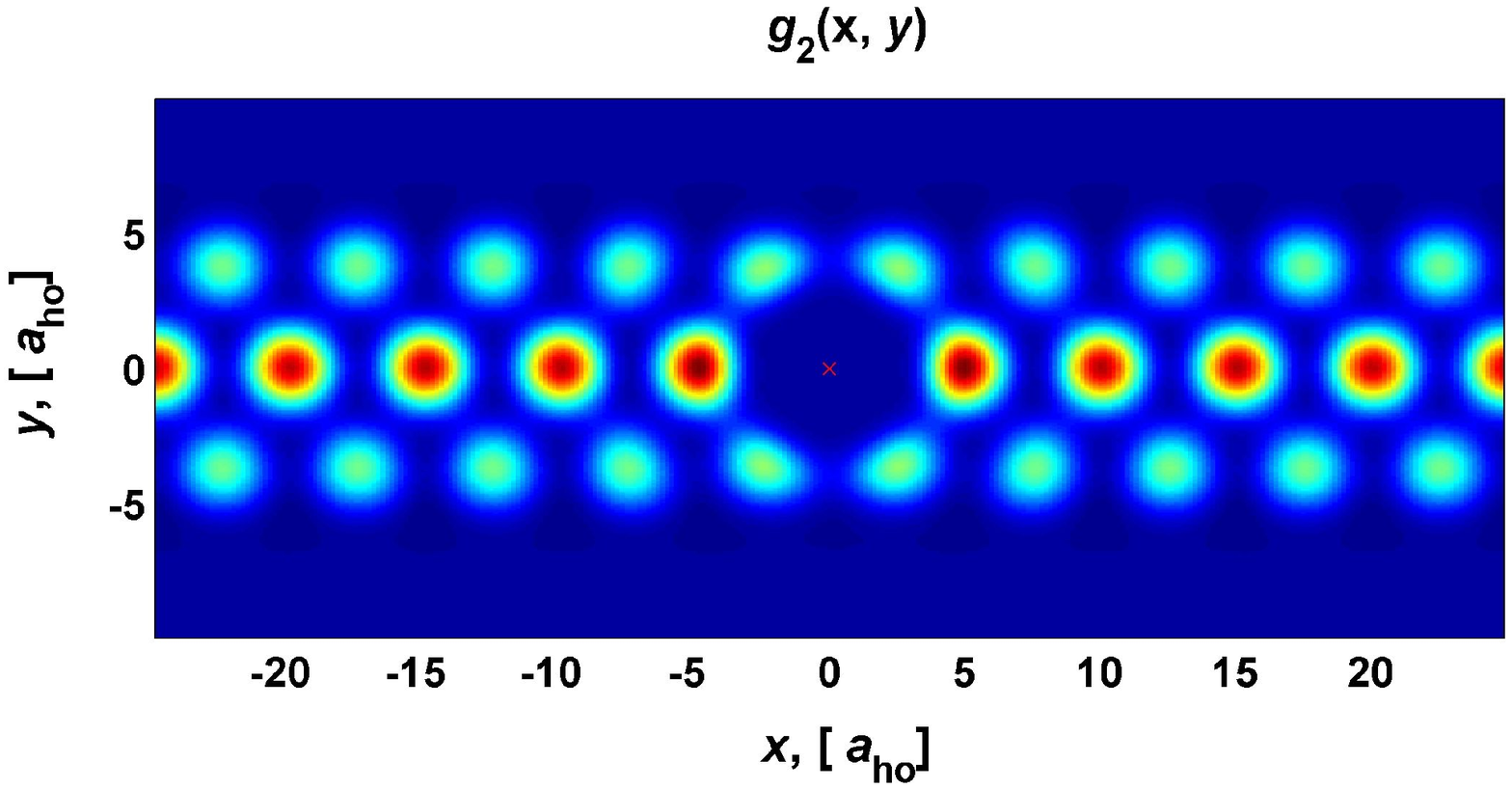}
\includegraphics[width=0.49\columnwidth]{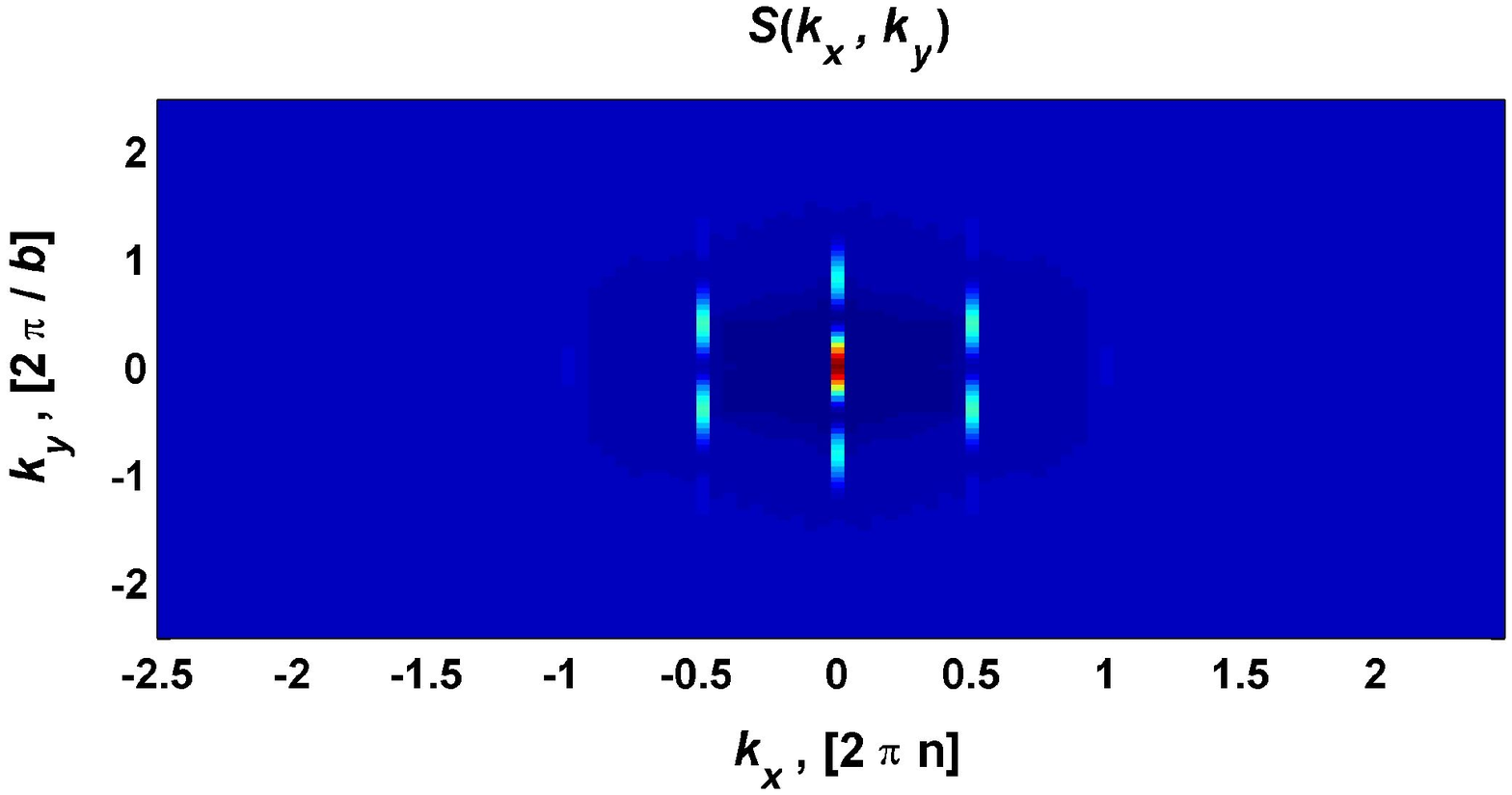}
\caption{(color online) Same as in Fig.~\protect\ref{fig:QNT}, but with $\tilde r_0=100$. The static structure form factor is displayed as in Fig.~\ref{fig:PDzigzag}.}
\label{fig:QNT:ZZ}
\end{center}
\end{figure}

Figure~\ref{fig:QNT:ZZ} displays the pair correlation function and the static structure factor of the same quantum system at zero temperature for the parameter regime, in which one observes a quasi-ordered zigzag structure in the classical system, compare with Fig.~\ref{fig:PDzigzag}. Now, quasi-ordering of the particle is visible in the pair correlation function, which corresponds to the appearance of well-defined peaks in the static structure form factor. Nevertheless, one does not observe secondary peaks, showing that there is still no true crystalline structure. Comparison with the classical case, Fig.~\ref{fig:PDzigzag}, shows similarities to the finite-temperature behavior in a regime when there is no ordering, while strong correlations are still present.

Let us now comment on the behavior of the static structure form factor in the quantum regime. In a dipolar quantum system the low-momentum part is described by phonons. The static structure factor grows linearly with $k$ for small momenta (see Fig.~\ref{fig:QNT}):
\begin{eqnarray}
S(k) = \frac{\hbar |k|}{2mc},\quad k\to 0,
\label{eq:Skphonons}
\end{eqnarray}
where $c$ is the speed of sound, see also Eq.~(\ref{eq:sound}). At zero temperature in the $k\to 0$ limit the static structure factor vanishes $S(k)\to 0$\footnote[2]{Note that from the definition (\ref{eq:Sk}) the value of the $S(k)$ is discontinuous at zero momentum where it gives the number of particles for $k=0$. For a gas $S(k)-N\delta(k)$ is instead continuous.}. The low-momentum behavior is important as it reflects the number fluctuations in a large system $\langle (\delta N)^2\rangle = N S(0)$ as it is discussed in details in \ref{App:B}, refer to Eq.~(\ref{eq:dN2:c}). On the other hand, the number fluctuations at finite temperature are related to the isothermal compressibility\cite{TheoreticalPhysicsV} $c^2_T = 
- V^2/(mN)\;(\partial P/\partial V)_T$ 
as
\begin{eqnarray}
\langle (\delta N)^2\rangle
= -\frac{k_BT N^2}{V^2} \left(\frac{\partial V}{\partial P}\right)_{T}
= \frac{Nk_BT}{mc_T^2}
\end{eqnarray}
This means that at finite temperatures $S(k)$ no longer vanishes in the zero momentum limit, instead it goes to a constant value $S(k) \to k_BT /mc_T^2, k\to 0$, defined by the temperature and the compressibility (\ref{eq:sound}).
Moreover, the effects of the temperature in quantum Bose systems on the low-momentum part of the static structure factor can be found analytically \cite{TheoreticalPhysicsIX} 
and $S(k)$ is given by the following expression
\begin{eqnarray}
S(k) = \frac{\hbar k}{2mc}\cth\frac{\hbar kc}{2k_BT}
= \frac{k_BT}{mc^2} + \frac{\hbar^2k^2}{12m k_B T} + ...,\quad |k|\ll k_BT/\hbar c
\label{eq:SkQNTthermal}
\end{eqnarray}
The low-momenta dependence is no longer linear, but a quadratic one with the coefficient of proportionality entirely defined by the temperature.

The one-dimensional case has been studied in the literature~\cite{Citro07,Citro08,Mazzanti08}, and evidence has been given that the one-dimensional dipolar systems manifest Luttinger liquid behavior even if the interactions are of a long-range type~\cite{Citro07}. From Haldane's approach \cite{Haldane81} it is possible to write asymptotic expressions for the one-dimensional pair correlation function $g_2(x)$. Although this effective description is valid only at large distances, it is sufficient to predict the behavior of the peaks in the static structure factor using properties of the Fourier transformation~(\ref{eq:Sk}). The static structure factor has been computed for various values of $\tilde n\tilde r_0$ in Refs.~\cite{Citro08,Mazzanti08}, showing that at large values multiple peaks appear. Their height decays with the peak order according to a power law decay (with the exponent depending on the Luttinger parameter, and hence on the linear density~\cite{Mazzanti08}), showing interesting analogies with the classical case at finite temperatures. Clear important differences between quantum mechanical and thermal effects in the structure form factor $S_k$ can be identified in its behavior at low values of $k$, as the value of the static structure factor at zero momentum is related to the integral of the density fluctuations over the whole volume of the system. In the quantum mechanical case, it grows from zero linearly in $k$ (see Fig.~\ref{fig:QNT}) while in the classical case at finite $T$ no longer vanishes at zero momentum, instead it goes to a constant value, defined by the temperature and compressibility (see Appendix B for details).  This can be lead back to the issue how number fluctuations $\langle (\delta N)^2\rangle $ increase with the volume size $L^D$, $D$ being dimensionality of the homogeneous space. Indeed, thermal fluctuations increase linearly with the volume, while fluctuations in quantum systems at zero temperature follow the law $\langle (\delta N)^2\rangle \propto L^{D-1}\ln C L$ \cite{Giorgini98,Astrakharchik:063616}, where $C$ is a constant. In particular, in a one-dimensional system (and also in a quasi-one-dimensional geometry), the fluctuations follow the logarithmic dependence $\langle (\delta N)^2\rangle \propto \ln L$ \cite{Levitov02eng}.

\section{Conclusions} \label{Sec:8}

We have discussed the properties of the classical and quantum distribution of a low-dimensional gas of dipolar particles at thermal equilibrium, by computing numerically the pair correlation function and the static structure form factor, focusing to the regime where one would expect quasi-ordered structures of dipoles in a linear or zigzag structure. Thermal and quantum fluctuations tend to destroy ordering, bringing further evidence that a true self-organized crystal cannot exist at finite temperature or (in a quantum system) at a finite interaction strength. Analogies and differences between the effects of thermal and quantum fluctuations on the phase diagram, the pair correlation function and the static structure factor are discussed. Behavior of the static structure factor for small momenta is addressed. In the limit of large dipolar interaction strengths or of large linear density, the transition from a one-dimensional to a planar configuration occurs with the creation of a mesoscopic structure in the transverse direction, which exhibits the main features of the transition from a single to a double and then a multiple chain distribution of dipoles, while quasi-order is observed also in the transverse pair correlation function at large densities. Such patterns are characterized by non-local correlations, which arise from zero-point fluctuations, and which may be important resources for the realization of quantum simulators~\cite{Maciej-Review,Zoller08}.

\vspace{0.5cm}
Support by the ESF (EUROQUAM "CMMC"), the European Commission
(EMALI, MRTN-CT-2006-035369; SCALA, Contract No.\ 015714) and the
Spanish Ministerio de Educaci\'on y Ciencia (Consolider Ingenio
2010 ``QOIT''; FIS2007-66944;  FIS2008-04403; Ramon-y-Cajal; Juan de la Cierva) are
acknowledged.

\begin{appendix}

\section{Transition from linear chain to the zigzag structure}
\label{App:A}

For very large interaction strengths,  the linear chain becomes unstable for $\tilde n > \tilde n_c$. Under this condition one can show that, within a certain interval of densities, the stable configuration is a planar zigzag structure. The zigzag structure is composed by two chains, one shifted with respect to the other by a lattice distance $a$, so that with respect to the linear chain the periodicity doubles from $a$ to $2a$. Following~\cite{Fishman08}, we here make the Landau theory of the classical phase transition from a linear chain of dipoles to a zigzag, which is strictly valid when thermal effects can be neglected.

We identify the order parameter of the zigzag as the distance $b/2$ of the dipoles from the trap axis $x$ and the control parameter as the density $\tilde n$ while the transverse trap frequency $\nu_t$ remains fixed. A similar analysis can be performed using the transverse frequency as the control parameter and fixing the density, see~\cite{Fishman08}. In order to study the behavior close to the instability point, we expand Eq.~\eref{eq:tildeV} up to the fourth order around the equilibrium positions of the chain, $\tilde V=\sum_{\ell=0}^4\tilde V^{(\ell)}$, where $\ell$ labels the order. Here, $\tilde V^{(0)}$ is the ground state energy in the limit of strong correlations, which for $N\gg 1$ reads
\begin{equation}
\label{eq:groundstateenergy}
\tilde V^{(0)} = N \frac{\tilde r_0 \zeta(3)}{a^3}.
\end{equation}
Moreover, as we are expanding close to the equilibrium positions of the linear chain, the first order term $V^{(1)}$ vanishes. Using the decomposition into the eigenmodes of the linear chain, the quadratic term takes the form \begin{equation} \tilde V^{(2)}=\frac{1}{2}\sum_{k}\sum_{s=\pm}\left(\omega_\parallel(k)^2\Theta_k^{(s)2}+\alpha(k)\Psi^{(s)2}_k\right)
\end{equation} where \begin{equation} \label{alpha:k} \alpha(k)=1-\frac{12 \tilde r_0}{a^5}\sum_{i>0}\frac{1}{i^5}\sin^2\frac{kia}{2} \end{equation}
and it coincides with $\tilde\omega_{\perp}(k)^2$, Eq.~\eref{Eq:omega_perp}, for $\tilde n< \tilde n_c$, while the normal modes $\Theta_k^{(s)}$ and $\Psi_k^{(s)}$ are related to axial and transverse displacements by the relation
\begin{eqnarray}
 \tilde q_j&=&\sqrt{\frac{2}{N}} \sum_{k} \left(\Theta_k^{(+)}\cos kja+\Theta_k^{(-)}\sin kja\right)
 \label{Fourier:q} \\
 \tilde w_j&=&\sqrt{\frac{2}{N}}\sum_{k}\left(\Psi_k^{(+)}\cos kja+\Psi_k^{(-)}\sin kja\right)
 \label{Fourier:w}
\end{eqnarray}
with $k=2 \pi l /Na$ with $l=0,1, 2,\ldots,N/2$ and the subscript $\pm$ indicates the parity by the reflection $k\to -k$. For convenience, we denote by $\Psi_0$ the mode of the linear chain at wave vector $k_0\equiv\pi/a$ (zigzag mode). The linear chain becomes mechanically unstable when, by varying $\tilde n$, the frequency of the zigzag mode with wave vector $k=\pi/a$, Eq.~(\ref{soft:mode}), vanishes. Around the instability point the zigzag mode is significantly coupled to other quasi-degenerate modes by the third and fourth order terms $\tilde V^{(3)}$ and $\tilde V^{(4)}$. These quasi-degenerate modes are long wavelengths axial modes $\Theta_{\delta k}$ at wave vectors $\delta k$, such that $|\delta k|a\ll 1$, and short wavelength transverse modes $\Psi_{k_0+\delta k}$ at wave vector $k=k_0+\delta k$, with $|\delta k|a\ll 1$. At lowest order in the expansion in $\delta k a$, the effective potential for the modes which are relevantly coupled at the instability reads \begin{eqnarray} \label{V:eff}
\tilde V_{\rm eff}=\frac{1}{2}\alpha_{0}
\Psi^2_{0}+\frac{1}{2}\sum_{\delta k> 0}\alpha_{\delta
k}\left(\Psi_{\delta k}^{(+)2}+\Psi_{\delta k}^{(-)2}\right)
+\tilde V^{(4)}_{k\simeq k_0} \end{eqnarray} where for brevity we denote $\alpha_{\delta k}\equiv\alpha(k_0-\delta k)$ and $\Psi_{\delta k} \equiv\Psi_{k_0+\delta k} $. The fourth order potential for the relevant modes is given by: \begin{eqnarray}
\tilde V^{(4)}_{k\simeq k_0}= A\Psi_{0}^4 +6
A\Psi_{0}^2\sum_{\delta k>0} \left({\Psi_{\delta
k}^{(+)}}^2+{\Psi_{\delta k}^{(-)}}^2\right) + A
\Psi_{0}f\left(\Psi^{(+)}_{\delta k_1},\Psi_{\delta
k_2}^{(-)},\Psi_{\delta k_1+\delta k_2}^{(-)}\right) +O(\delta k^2
a^2) \end{eqnarray} where \begin{equation} \label{eq:A} A=\frac{30
\tilde r_0 \tilde n_c^7}{N}\left(1-\frac{1}{2^7}\right)\zeta(7)
\end{equation}
and the function $f$ contains a sum of products of three amplitudes $\Psi_{\delta k}$ for $\delta k \neq 0$,
whose precise form is not important for the present discussion. We now allow the linear density $\tilde n$ to take values in the interval $[\tilde n_c\!-\!\delta,\tilde n_c\!+\!\delta]$, such that $\alpha_{\delta k}$ may take only small but negative values, and for $\alpha_{\delta k}<0$ we determine the corrections $\bar{\Psi}_{\delta k}^{(\pm)}$ to the equilibrium positions of the linear chain using Eq.~(\ref{V:eff}), assuming that the role of the fourth order potential $\tilde V^{(4)}_{k\simeq k_0}$ is to give rise to a small displacement $b\ll a$ with respect to the equilibrium interparticle distance $a$. Using a stability analysis of the zigzag configuration in a similar way as the one used for the linear--zigzag transition in low dimensional ion crystals \cite{Fishman08}, we find that the stable solution has $\bar{\Psi}_{\delta k}=0$ for $\delta k>0$ and
\begin{equation}
\bar{\Psi}_0=\left(-\frac{\alpha_0}{4A}\right)^{1/2}
\label{eq:solrho} \end{equation} where $\alpha_0=1-(\tilde n/\tilde n_c )^5<0$. Correspondingly, the spatial displacement $b$ reads $b=2\bar{\Psi}_0/\sqrt{N}$ and, using Eqs.~\eref{eq:nc} and Eq.~\eref{eq:A} it can be written as \begin{equation} \label{Eq:b:Landau} b=\mathcal B a
\sqrt{1-\left(\frac{\tilde n_c}{\tilde n}\right)^5} \end{equation}
where we have introduced the dimensionless constant
\begin{equation}
\mathcal
B=\sqrt{\frac{8}{5}\frac{31}{127}\frac{\zeta(5)}{\zeta(7)}} =
0.634...
\end{equation}
Note that the expression for $b$ found in Eq.~(\ref{Eq:b:Landau}) corresponds to the second order expansion of Eq.~(\ref{Eq:b:exact}) in the parameter $b/a$. The exact result found by solving numerically Eq.~(\ref{Eq:b:exact}) and the approximated result Eq.~(\ref{Eq:b:Landau}) are compared in Fig.~\ref{fig:b}.

The jump in the classical ground-state energy at the transition point is given by the difference between the potential of the soft mode at each sides of the critical points. We denote by
\begin{eqnarray} \label{V:soft} \tilde V^{\rm soft}
=\frac{1}{2}\alpha_0 \Psi_0^2+A\Psi_0^4, \end{eqnarray} the effective potential for the soft mode, where $A$ is given by Eq.~\eref{eq:A}. At zero temperature, for $\tilde n= \tilde n_c-\delta$, with $\delta>0$ then $\tilde V^{\rm soft}= 0$, while for $\tilde n= \tilde n_c+\delta$ then $\tilde V^{\rm soft}\neq 0$. Letting $\delta\to 0^+$ we find that the energy difference per particle is given by \begin{equation} \Delta \tilde E=\frac{\tilde
V^{\rm soft}(\tilde n\to \tilde n_c^+)-\tilde V^{\rm soft}(\tilde
n\to \tilde n_c^-)}{N} =-\frac{1}{2}\mathcal C (\tilde n_c-\tilde
n)^2 \end{equation}
where $\mathcal C= [4650 \zeta(5)]/[127 \zeta(7)]=37.65\ldots$. Notice that the second derivative of $\Delta \tilde E$ with respect  to $\tilde n$ is discontinuous at the critical point. At zero temperature this means that the phase transition is of the second order.

\section{Static structure form factor and thermodynamic properties \label{App:B}}

Small momentum behavior is directly related to the density fluctuations in the system. Indeed, the number fluctuations in volume $L$ is related to the integral of pair correlation function
\begin{eqnarray}
\langle (\delta N)^2\rangle
= \int_L\!dx_1\!\int_L\!dx_2\; \langle \delta n(x_1)\;\delta n(x_2)\rangle
= N+\int_L\!dx_1\!\int_L\!dx_2\; [g_2(x_1\!-\!x_2)-n^2]
\label{eq:dN2}
\end{eqnarray}
Here we consider the case of a linear chain considering it as a one-dimensional objects so that the volume is linear $L$ and the pair correlation function saturates to a constant value, $g_2(x)\to n^2$ as $|x|\to\infty$. The static structure factor is related by the Fourier transformation to $g_2(x)$ according to relation $g_2(x) = n^2 + n\int(S(k)-1) e^{-ikx}\;dk/2\pi$, which is inverse to Eq.~(\ref{eq:Sk}). By substituting this expression to Eq.~(\ref{eq:dN2}) one gets
\begin{eqnarray}
\langle (\delta N)^2\rangle
= N+n \int_L\!dx_1\!\int_L\!dx_2\!\int_{-\infty}^\infty \frac{dk}{2\pi}(S(k)-1) e^{-ikx}
\label{eq:dN2:b}
\end{eqnarray}
Changing to center of mass variables $x=x_1-x_2$; $X=(x_1+x_2)/2$ one separates spatial integrals into two. One is trivial $\int_L dX = L$. The second gives the $\delta$-function $\int e^{ikx}dx=2\pi\delta(k)$ leading to a simple relation
\begin{eqnarray}
\langle (\delta N)^2\rangle = N S(0).
\label{eq:dN2:c}
\end{eqnarray}
\end{appendix}

\newpage
\section*{Bibliography}


\begin{thebibliography}{10}

\bibitem{Dubin99}
Daniel H.~E. Dubin and T.~M. O\char39{}Neil.
\newblock Trapped nonneutral plasmas, liquids, and crystals (the thermal
  equilibrium states).
\newblock {\em Rev. Mod. Phys.}, 71(1):87, Jan 1999.

\bibitem{Birkl92}
G.~Birkl, S.~Kassner, and H.~Walther.
\newblock Multiple-shell structures of laser-cooled $^{24}$Mg$^{+}$ ions in a
  quadrupole storage ring.
\newblock {\em Nature}, 357:310, 1992.

\bibitem{Waki92}
I.~Waki, S.~Kassner, G.~Birkl, and H.~Walther.
\newblock Observation of ordered structures of laser-cooled ions in a
  quadrupole storage ring.
\newblock {\em Phys. Rev. Lett.}, 68(13):2007--2010, Mar 1992.

\bibitem{Drewsen08}
M.~Drewsen, A.~Mortensen, E.~Nielsen, and T.~Matthey.
\newblock Strongly correlated ion Coulomb systems.
\newblock In A.~Campa, A.~Giansanti, G.~Morigi, and F.~Sylos-Labini, editors,
  {\em (AIP Proceedings, Melville, 2008)}, volume 970 of {\em Dynamics and
  thermodynamics of systems with long-range interactions: Theory and
  Experiments}, pages 295--302, 2008.

\bibitem{Javanainen95}
Jiaju Yin and Juha Javanainen.
\newblock Quantum motion of two trapped ions in one dimension.
\newblock {\em Phys. Rev. A}, 51(5):3959--3966, May 1995.

\bibitem{Retzker08}
A.~Retzker, R.~C. Thompson, D.~M. Segal, and M.~B. Plenio.
\newblock Double well potentials and quantum phase transitions in ion traps.
\newblock {\em Phys. Rev. Lett.}, 101(26):260504, 2008.

\bibitem{Pfau07}
Thierry Lahaye, Tobias Koch, Bernd Fr\"ohlich, Marco Fattori, Jonas Metz, Axel
  Griesmaier, Stefano Giovanazzi, and Tilman Pfau.
\newblock Strong dipolar effects in a quantum ferrofluid.
\newblock {\em Nature}, 448:672--675, 2007.

\bibitem{Pfau08}
T.~Koch, T.~Lahaye, J.~Metz, B.~Fr\"ohlich, A.~Griesmaier, and T.~Pfau.
\newblock Stabilizing a purely dipolar quantum gas against collapse.
\newblock {\em Nature Physics}, 4:218--222, 2008.

\bibitem{buchler07}
H.~P. Buchler, E.~Demler, M.~Lukin, A.~Micheli, N.~Prokof'ev, G.~Pupillo, and
  P.~Zoller.
\newblock Strongly correlated 2d quantum phases with cold polar molecules:
  controlling the shape of the interaction potential.
\newblock {\em Phys. Rev. Lett.}, 98:060404, 2007.

\bibitem{Astrakharchik07a}
G.~E. Astrakharchik, J.~Boronat, I.~L. Kurbakov, and Yu.~E. Lozovik.
\newblock Quantum phase transition in a two-dimensional system of dipoles.
\newblock {\em Phys. Rev. Lett.}, 98(6):060405, 2007.

\bibitem{Goral02}
K.~G\'oral, L.~Santos, and M.~Lewenstein.
\newblock Quantum phases of dipolar bosons in optical lattices.
\newblock {\em Phys. Rev. Lett.}, 88:170406, 2002.

\bibitem{Santos03}
L.~Santos, G.~V. Shlyapnikov, and M.~Lewenstein.
\newblock Roton-maxon spectrum and stability of trapped dipolar Bose-Einstein
  condensates.
\newblock {\em Phys. Rev. Lett.}, 90(25):250403, Jun 2003.

\bibitem{Menotti07}
C.~Menotti, C.~Trefzger, and M.~Lewenstein.
\newblock Metastable states of a gas of dipolar bosons in a 2d optical lattice.
\newblock {\em Phys. Rev. Lett.}, 98(23):235301, 2007.

\bibitem{Kollath08}
C.~Kollath, Julia~S. Meyer, and T.~Giamarchi.
\newblock Dipolar bosons in a planar array of one-dimensional tubes.
\newblock {\em Phys. Rev. Lett.}, 100(13):130403, 2008.

\bibitem{Parker08}
N.~G. Parker and D.~H.~J. O'Dell.
\newblock Thomas-fermi versus one- and two-dimensional regimes of a trapped
  dipolar Bose-Einstein condensate.
\newblock {\em Phys. Rev. A}, 78(4):041601, 2008.

\bibitem{Nath09}
R.~Nath, P.~Pedri, and L.~Santos.
\newblock Phonon instability with respect to soliton formation in
  two-dimensional dipolar Bose-Einstein condensates.
\newblock {\em Phys. Rev. Lett.}, 102(5):050401, 2009.

\bibitem{Kinoshita05}
Toshiya Kinoshita, Trevor Wenger, and David~S. Weiss.
\newblock Local pair correlations in one-dimensional Bose gases.
\newblock {\em Phys. Rev. Lett.}, 95(19):190406, 2005.

\bibitem{Citro07}
R.~Citro, E.~Orignac, S.~De Palo, and M.~L. Chiofalo.
\newblock Evidence of luttinger-liquid behavior in one-dimensional dipolar
  quantum gases.
\newblock {\em Phys. Rev. A}, 75(5):051602, 2007.

\bibitem{Arkhipov05}
A.~S. Arkhipov, G.~E. Astrakharchik, A.~V. Belikov, and {\relax Yu}~E. Lozovik.
\newblock Ground-state properties of a one-dimensional system of dipoles.
\newblock {\em JETP Lett.}, 82:39, 2005.

\bibitem{Astrakharchik08b}
G.~E. Astrakharchik, Giovanna Morigi, Gabriele {\relax De Chiara}, and
  J.~Boronat.
\newblock Ground state of low-dimensional dipolar gases: Linear and zigzag
  chains.
\newblock {\em Phys. Rev. A}, 78(6):063622, 2008.

\bibitem{Fishman08}
S.~Fishman, G.~{\relax De Chiara}, T.~Calarco, and G.~Morigi.
\newblock Structural phase transitions in low-dimensional ion crystals.
\newblock {\em Phys. Rev. B}, 77:064111, 2008.

\bibitem{Rabl}
P.~Rabl and P.~Zoller.
\newblock Molecular dipolar crystals as high-fidelity quantum memory for hybrid
  quantum computing.
\newblock {\em Phys. Rev. A}, 76(4):042308, 2007.

\bibitem{Ashcroft}
Neil~W. Ashcroft and David~N. Mermin.
\newblock {\em Solid State Physics}.
\newblock Thomson Learning, Toronto, January 1976.

\bibitem{Pitaevskii03}
L.~P. Pitaevskii and S.~Stringari.
\newblock {\em Bose-Einstein Condensation}.
\newblock Oxford University Press, Oxford, 2003.

\bibitem{Boronat94b}
J.~Boronat and J.~Casulleras.
\newblock Monte Carlo analysis of an interatomic potential for he.
\newblock {\em Phys. Rev. B}, 49:8920, 1994.

\bibitem{Sarsa02}
A.~Sarsa, J.~Boronat, and J.~Casulleras.
\newblock Quadratic diffusion Monte Carlo and pure estimators for atoms.
\newblock {\em J. Chem. Phys}, 116:5956, 2002.

\bibitem{Girardeau60}
M.~Girardeau.
\newblock Relationship between systems of impenetrable bosons and fermions in
  one dimension.
\newblock {\em J. Math. Phys. (N.Y.)}, 1:516, 1960.

\bibitem{Citro08}
S.~De Palo, E.~Orignac, R.~Citro, and M.~L. Chiofalo.
\newblock Low-energy excitation spectrum of one-dimensional dipolar quantum
  gases.
\newblock {\em Physical Review B (Condensed Matter and Materials Physics)},
  77(21):212101, 2008.

\bibitem{TheoreticalPhysicsV}
L.~D. Landau and E.~M. Lifshitz.
\newblock {\em Statistical Physics, Part 1}.
\newblock Pergamon Press, Oxford, 1980.

\bibitem{TheoreticalPhysicsIX}
E.~M. Lifshitz and L.~P. Pitaevskii.
\newblock {\em Statistical Physics, Part 2}.
\newblock Pergamon Press, Oxford, 1980.

\bibitem{Mazzanti08}
F.~Mazzanti, G.~E. Astrakharchik, J.~Boronat, and J.~Casulleras.
\newblock Ground-state properties of a one-dimensional system of hard rods.
\newblock {\em Phys. Rev. Lett.}, 100(2):020401, 2008.

\bibitem{Haldane81}
F.~D.~M. Haldane.
\newblock Effective harmonic-fluid approach to low-energy properties of
  one-dimensional quantum fluids.
\newblock {\em Phys. Rev. Lett.}, 47:1840, 1981.

\bibitem{Giorgini98}
S.~Giorgini, L.~P. Pitaevskii, and S.~Stringari.
\newblock Anomalous fluctuations of the condensate in interacting Bose gases.
\newblock {\em Phys. Rev. Lett.}, 80(23):5040--5043, Jun 1998.

\bibitem{Astrakharchik:063616}
G.~E. Astrakharchik, R.~Combescot, and L.~P. Pitaevskii.
\newblock Fluctuations of the number of particles within a given volume in cold
  quantum gases.
\newblock {\em Phys. Rev. A}, 76(6):063616, 2007.

\bibitem{Levitov02eng}
L.~S. Levitov and A.~V.Shitov.
\newblock {\em Green's functions. Problems and solutions. 2nd edit. [in
  Russian]}.
\newblock Fizmatlit, 2002.

\bibitem{Maciej-Review}
Maciej Lewenstein, Anna Sanpera, Veronica Ahufinger, Bogdan Damski, Aditi Sen,
  and Ujjwal Sen.
\newblock Ultracold atomic gases in optical lattices: mimicking condensed
  matter physics and beyond.
\newblock {\em Advances in Physics}, 56:243, 2007.

\bibitem{Zoller08}
G.~Pupillo, A.~Griessner, A.~Micheli, M.~Ortner, D.-W. Wang, and P.~Zoller.
\newblock Cold atoms and molecules in self-assembled dipolar lattices.
\newblock {\em Phys. Rev. Lett.}, 100(5):050402, 2008.

\end{thebibliography}

\end{document}